\documentstyle[12pt,emulateapj,psfig]{article}

\newcommand{\gtilde}
 {~ \raisebox{-1ex}{$\stackrel{\textstyle >}{\sim}$} ~}
\newcommand{\ltilde}
 {~ \raisebox{-1ex}{$\stackrel{\textstyle <}{\sim}$} ~}
\def\ltsima{$\; \buildrel < \over \sim \;$}
\def\ltsim{\lower.5ex\hbox{\ltsima}}
\def\gtsima{$\; \buildrel > \over \sim \;$}
\def\gtsim{\lower.5ex\hbox{\gtsima}}

\begin{document}
\small
\footnotesize

\submitted{To Appear in the Astrophysical Journal}

\title{Unavoidable Selection Effects in the Analysis of Faint 
Galaxies in the Hubble Deep Field: Probing the Cosmology and 
Merger History of Galaxies}

\author{Tomonori Totani$^1$ and Yuzuru Yoshii$^{2,3}$}
\altaffiltext{1}{ 
National Astronomical Observatory, Mitaka, Tokyo 181-8588,
Japan (E-mail: totani@th.nao.ac.jp)}
\altaffiltext{2}{Institute of Astronomy, School of Science,
The University of Tokyo, 2-21-1 Osawa, Mitaka, Tokyo 181-8588, Japan}

\altaffiltext{3}{Research Center for the Early Universe, Faculty of Science,
The University of Tokyo, Tokyo 113-0033, Japan}

\date{\today}

\begin{abstract}
We present a detailed analysis of the number count and photometric 
redshift distribution of faint galaxies in the Hubble Deep Field 
(HDF), paying a special attention to the selection effects including
the cosmological dimming of surface brightness of galaxies, under the 
observational condition employed in this field.  We find a considerably 
different result from previous studies ignoring the selection effects, 
and these effects should therefore be taken into account in the analysis.
We find that the model of pure luminosity evolution (PLE) of galaxies 
in the Einstein-de Sitter (EdS) universe predicts much smaller counts 
than those observed at faint magnitude limits by a factor of more than 
10, so that a very strong number evolution of galaxies with $\eta \gtilde$ 
3--4 must be invoked to reproduce the $I_{814}$ counts, when parametrized 
as $\phi^* \propto (1+z)^\eta$.  However we show that such a strong number 
evolution under realistic merging processes of galaxies can not explain 
the steep slope of the $B_{450}$ and $V_{606}$ counts, and it is seriously 
inconsistent with their photometric redshift distribution.  We find that 
these difficulties still persist in an open universe with $\Omega_0 \gtilde 
0.2$, but are resolved only when we invoke a $\Lambda$-dominated flat 
universe, after examining various systematic uncertainties in modeling 
the formation and evolution of galaxies.  The present analysis revitalizes
the practice of using faint number counts as an important cosmological
test, giving one of the
arguments against the EdS universe and suggests acceleration 
of the cosmic expansion by vacuum energy density.  While a modest number 
evolution of galaxies with $\eta \ltilde 1$ is still necessary
even in a $\Lambda$-dominated universe, a stronger number evolution with 
$\eta > 1$ is rejected from the HDF data, giving a strong constraint on 
the merger history of galaxies.
\end{abstract}

\keywords{cosmology: observations --- 
galaxies: evolution --- galaxies: formation}

\section{Introduction}
Number counting of faint galaxies is one of the most fundamental 
observational tests with which the formation/evolution of galaxies as 
well as the geometry of the universe is probed. The best view to date 
of the optical sky to faint flux levels is given by the Hubble Deep Field 
(HDF, Williams et al. 1996), and it provides a valuable information to a wide 
range of studies on galaxies and cosmology.  A comprehensive study of the 
HDF galaxy counts has been performed by Pozzetti et al. (1998), and they 
found that a simple model of pure luminosity evolution (PLE), 
in which galaxies evolve passively due to star formation histories without 
mergers or number evolution, gives a reasonable fit to the HDF counts 
in all the four passbands of $U$, $B$, $V$, and $I$, when an open universe 
with $\Omega_0 = 0.1$ is assumed. 

The increase of the number of galaxies with their apparent magnitude 
was originally proposed as a measure of the geometry of the universe
(Sandage 1961), and considerable efforts have been made along this line 
(e.g., Yoshii \& Takahara 1988; Fukugita, 
Takahara, Yamashita \& Yoshii 1990; Yoshii \& Peterson 1991).  
However, the obtained constraints on 
cosmological parameters based on the PLE model have not been thought 
deterministic because of possible number evolution of galaxies by 
mergers.  Particularly, when the PLE model is used,
the Einstein-de Sitter (EdS) universe ($\Omega_0=1$) 
underpredicts the observed galaxy counts at faint magnitudes,
but a simple model of galaxy number evolution can 
reproduce the observed counts and save the EdS universe (Rocca-Volmerange \& 
Guiderdoni 1990; Pozzetti et al. 1996).  This degeneracy between the 
effects of galaxy evolution and cosmology has been a major problem when 
one uses the galaxy number count to determine the geometry of the universe.

The information of redshifts is able to break such a degeneracy, because 
luminous galaxies at great distance are distinguishable from dwarf galaxies 
in a local universe.  Although most of the HDF galaxies are too faint to 
measure the spectroscopic redshifts, several catalogs of their photometric 
redshifts have been published (Sawicki, Lin, \& Yee 1997; Wang, Bahcall, \& 
Turner 1998; Fern\'andez-Soto, Lanzetta, \& Yahil 1999).  The follow-up 
studies based on these catalogs show that the photometric redshifts give 
reasonably reliable estimates of spectroscopic redshifts and are useful for a 
statistical study of the HDF galaxies.  Here we give a combined analysis 
for the HDF counts and redshifts and constrain the cosmological parameters 
separately from the merger history of galaxies. 

Both the number count of faint galaxies and their redshift distribution 
are significantly affected by the selection effects inherent in the method
of detecting galaxies in faint surveys, but these important effects have
been ignored in almost all previous studies except for Yoshii \& Fukugita
(1991) and Yoshii (1993).  It is well known that the surface brightness 
of galaxies rapidly becomes dimmer with increasing redshift as $\propto 
(1+z)^{-4}$ (Tolman 1934), and this cosmological dimming makes many 
high-redshift galaxies
remain undetected below the threshold value of surface brightness adopted
in a galaxy survey (Pritchet \& Kline 1981; Tyson 1984; Ellis, 
Sievers, \& Perry 1984).  
The seeing or smoothing of an image furthermore lowers its
surface brightness, and the photometry scheme used in a survey heavily 
affects a magnitude estimate of the faintest galaxies.  Some observers 
apply corrections to raw counts of faint galaxies for those undetected, but 
it is in principle difficult and heavily model-dependent
to estimate the number of undetected galaxies.  
Rather, the best way is to make theoretical predictions with the selection 
effects taken into account and then compare them directly 
to raw counts (Yoshii 1993).

This paper is the first analysis of the HDF galaxies in which the 
above selection
bias against high-redshift galaxies is explicitly incorporated.  We use a 
standard PLE model of galaxies including the effects of internal dust 
obscuration and intergalactic HI absorption.  Number evolution of galaxies 
is also allowed for with simple modifications to the PLE model.  Throughout 
this paper, we use the AB photometry system with the notation of $U_{300}$, 
$B_{450}$, $V_{606}$, and $I_{814}$ (Williams et al. 1996).
In \S \ref{section:model}, we present a detailed description for
models of galaxy evolution and formulations to calculate galaxy
counts and redshift distribution with the selection effects taken into
account. Extensive calculations of number 
count predictions and comparison to the HDF counts are given in  \S
\ref{section:counts}, 
checking in great detail the uncertainties arising from the 
prescribed properties of local galaxies and their evolution.  
We will give the comparison of the model predictions with 
the observed photometric redshift distribution in \S \ref{section:redshifts}.
We discuss the results in \S \ref{section:discussion}.  The summary and 
conclusion of this paper are given in \S \ref{section:conclusions}.

\section{The Model of Galaxies and Detection in the HDF}
\label{section:model}
First we describe the basic ingredients involved in our theoretical modeling
such as the local luminosity function, galaxy evolution in luminosity and
number, internal and intergalactic absorption, and the selection effects.
Then we will present the formulations to calculate the number count of faint
galaxies and their redshift distribution.

\subsection{Galaxies at Present, and Their Evolution}
We use a standard PLE model of galaxy evolution, in which galaxies
are classified into five morphological types of E/S0, Sab, Sbc, Scd,
and Sdm. Spectral energy distributions (SEDs) and their evolution are
calculated by using the galaxy evolution model of Arimoto \& Yoshii 
(1987) for elliptical galaxies and the I1 model of
Arimoto, Yoshii, \& Takahara (1992) for spiral galaxies.
These models are constructed to reproduce the photometric and chemical 
properties of present-day galaxies. In order to see the systematic 
uncertainty in evolution models, we will also use an updated version of 
these models by Kobayashi et al. (1999) using the latest database of
stellar populations compiled by Kodama \& Arimoto (1997).  We set the 
epoch of galaxy formation at $z_F=5$ as a standard and change this value
to see the systematic uncertainty. 

The luminosity function of local galaxies is also important in  
predicting the number count and redshift distribution of faint
galaxies. We use the 
type-dependent (E/S0, Spiral, and Irr) $B$-band luminosity function 
derived from the Second Southern Sky Redshift Survey (SSRS2,
Marzke et al. 1998).  We associate the Sab, Sbc and Scd models to be 
assigned to spiral galaxies, whereas the Sdm model assigned to irregular 
galaxies.  The relative proportions of Sab, Sbc, and Scd are taken from
Pence (1976). In order to check the systematic uncertainty related to 
the luminosity function, we also use the type-independent luminosity 
function of Stromlo-APM redshift survey (Loveday et al. 1992) and
the type-dependent luminosity 
function from the Center for Astrophysics (CfA) redshift survey (Huchra 
et al. 1983).
Their Schechter parameters are tabulated in Table \ref{table:lf} (see also
Efstathiou,
Ellis, \& Peterson 1988 and Yoshii \& Takahara 1989).

\subsection{Absorption}
\label{section:absorption}
The above models of galaxy evolution do not include the absorption  
by interstellar dust which becomes significant for high-$z$ galaxies 
when observed in optical bands.  In order to take this effect into 
account, we make a physically natural assumption that the dust optical 
depth is proportional to the column density and metallicity of the 
gas.  In fact, it is well known that the Galactic extinction is 
well correlated to the column density of the HI gas (e.g., Burstein \& 
Heiles 1982). It is also known that the dust opacity becomes smaller 
in order of decreasing metallicity from the Galaxy to the Large and then
Small Magellanic Clouds, when the gas column density is fixed (e.g., 
Pei 1992).  Since the galaxy evolution models give the gas fraction 
$f_g$ and the metallicity $Z_g$ in the gas, the dust optical depth 
is calculated from $\tau_{\rm dust}=\kappa f_g Z_g r_e^{-2} (M/L_B)L_B$,  
where $r_e$, $M$, and $L_B$ are the effective radius, the baryon mass, and
the $B$-band luminosity of a galaxy, respectively.  (We will describe 
the treatment of galaxy size in \S \ref{section:selection}.)  
The proportionality 
constant $\kappa$ is chosen to be consistent with the present-day, 
average extinction of $A_V \sim 0.17$ taken from the Galactic extinction 
map (Burstein \& Heiles 1982; Schlegel, Finkbeiner, \& Davis 1998)
and a theoretical estimate (Hatano, Branch, \& Deaton 1998).
The standard extinction curve of our Galaxy (e.g., Pei 1992)
is used for the wavelength dependence of the optical depth.

Given the optical depth, the attenuation of emerging stellar lights 
depends on the spatial dust distribution.  Following Disney, Davies, 
\& Phillipps (1989), there are two extreme cases such as the screen 
model in which the dust is distributed on the line of sight to stars, 
and the slab model in which the dust has the same distribution with 
stars.  Neglecting the scattering of lights by dust, the attenuation 
factor of stellar lights is given by $\exp (-\tau_{\rm dust})$ for
the screen model and $[1-\exp (-\tau_{\rm dust})]/\tau_{\rm dust}$ for 
the slab model. In fact, the galaxy evolution model used here has been
made to reproduce the present-day SED of galaxies which has been already 
affected by dust obscuration. We take into account this point and
hence correct the above attenuation factors by using the optical depth
at present. Therefore the present-day SEDs of model galaxies are the
same for all the prescriptions of dust-free, screen, and slab models.

In the slab model the apparent reddening reaches an asymptote when 
the optical depth becomes much larger than unity, because the observed 
lights are emitted from surface regions of a galaxy where the optical 
depth to an observer is low.  However, the observed correlation between 
the power-law index of UV spectra and the Balmer line ratio, both of 
which are a reddening indicator, extends well beyond the asymptote. 
This indicates that the observed reddening of starburst galaxies is 
larger than expected from the slab model, and at least some fraction 
of dust should behave like a screen (for detail see Calzetti, Kinney, 
\& Storchi-Bergmann 1994).  

We then use the screen model as a standard, considering that UV and 
optical observations of starburst galaxies favor the screen dust.
It may be an extreme prescription that all dust is distributed as a 
screen, but note that the emergent lights are rapidly attenuated 
exponentially once a considerable fraction of dust contributes to the
screen.  Therefore the screen model is more appropriate than the slab 
model in which the attenuation factor decreases only moderately like 
$\tau_{\rm dust}^{-1}$ when $\tau_{\rm dust}\rightarrow \infty$.   

\begin{figure}
  \begin{center}
    \leavevmode\psfig{file=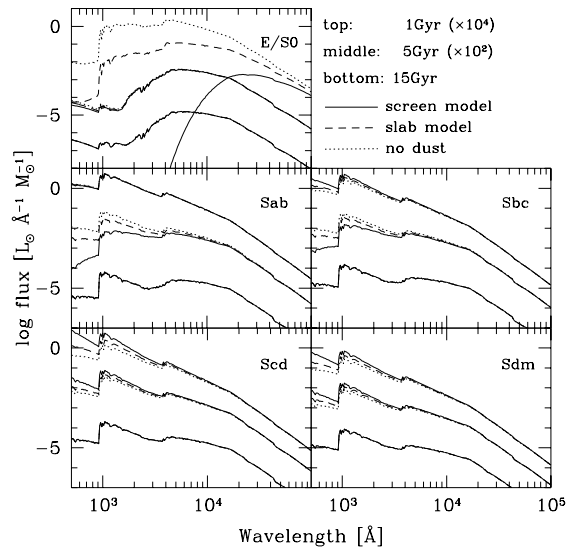,width=7.6cm}
  \end{center}
\caption{Spectral energy distributions (SEDs) of various morphological
types of galaxies at epochs of 1, 5, and 15 Gyrs (from the top to the bottom) 
after their formation. For the purpose of clarity, 
the SEDs at 1 and 5 Gyrs are artificially multiplied by a
factor of $10^4$ and $10^2$, respectively. 
The screen model of dust extinction is used as our standard and is 
incorporated in the calculations shown by the solid line. The dashed and
dotted lines are those from the slab dust model and the no-extinction
model, respectively.  The SEDs at 15 Gyrs are made to agree with the 
observed SEDs in the present-day galaxies.
}
\label{fig:sed-dust}
\end{figure}

The spectral energy distributions (SEDs) of galaxies of various types
are given in Fig. \ref{fig:sed-dust} at several epochs of galaxy 
evolution for three cases of dust extinction such as the screen model 
(solid line), the slab model (dashed line), and the no-extinction 
model (dotted line).
The effect of dust extinction is especially important for elliptical 
galaxies at high redshifts, where UV radiation is quite strong because 
of the initial starbursts supposed in the galactic wind model of 
elliptical galaxies (Arimoto \& Yoshii 1987).  According to the method
described above, the dust optical depth during the initial starburst
phase of elliptical galaxies is estimated to be much larger than unity 
($\tau \gtilde 10$) for UV photons at $\sim$ 2000 \AA \ 
in the restframe. 
On the other hand, the evolution of dust obscuration makes
UV luminosity of late-type spiral galaxies
brighter at early epochs than that without 
the dust effect, because of the lower
metal abundance than the present-day galaxies. However, this effect
in late-type spiral galaxies
is not significant because they are not heavily
obscured at present.

\begin{figure}
  \begin{center}
    \leavevmode\psfig{file=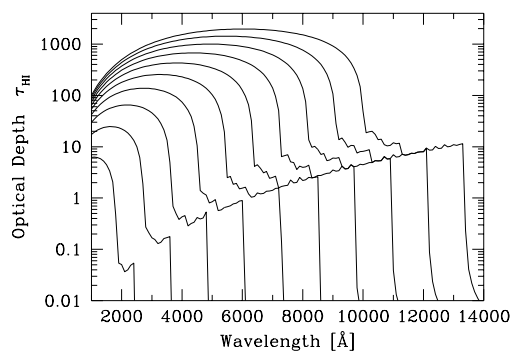,width=7.6cm}
  \end{center}
\caption{Optical depth of intergalactic absorption by intervening
HI clouds as a function of observed wavelength for various
source redshifts of $z$ = 1, 2, 3,..., 10 (from left to right).
Shown are the calculations by Yoshii \& Peterson (1994), with the
Doppler $b$-parameter of $b$ = 20 km/s in the HI clouds (Spitzer 
1978). 
}
\label{fig:h1-abs}
\end{figure}

In addition to the internal absorption by dust, the intergalactic 
absorption significantly affects the visibility of high-redshift galaxies
(Yoshii \& Peterson 1994; Madau 1995).  Lights from a distant galaxy 
at rest wavelengths below the Lyman limit (912 \AA) and those below the 
Lyman $\alpha$ line (1216 \AA) are extinguished by Lyman continuum 
absorption and Lyman series line absorption, respectively, in 
intergalactic HI 
clouds along the line of sight.  We include this effect consistently in 
our theoretical calculations making use of the intergalactic optical 
depth calculated by 
Yoshii \& Peterson (1994). The optical depth of this absorption
is shown in Fig. \ref{fig:h1-abs},
as a function of observed wavelength for various source redshifts.

\subsection{Selection Effects}
\label{section:selection}
Apparent surface brightness and size of an image in a survey 
observation are the essential quantities for it to be detected as a 
real galaxy.  We calculate these quantities of a model galaxy assuming 
its intrinsic luminosity profile and size and taking into account the 
cosmological dimming and the observational seeing.  For the details of 
the formulations, see Yoshii (1993).

\subsubsection{Galaxy Sizes and Luminosity Profile}
\label{section:size-luminosity}
\begin{figure}
  \begin{center}
    \leavevmode\psfig{file=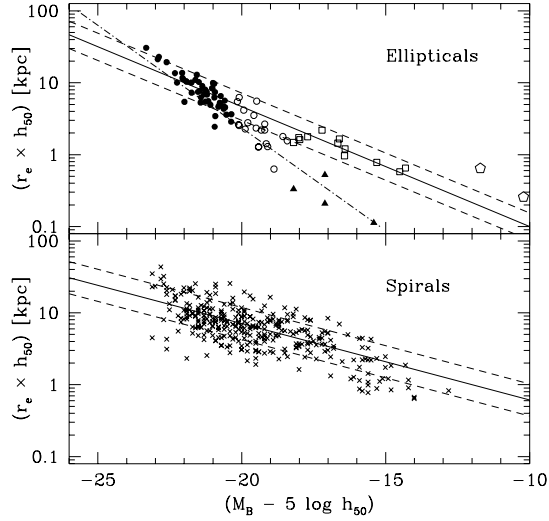,width=7.3cm}
  \end{center}
\caption{
Size-luminosity relations for local elliptical and spiral galaxies.
The data are from Bender et al. (1992) for elliptical galaxies
and Impey et al. (1996) for spiral galaxies.
For elliptical galaxies, the filled circles are for giant ellipticals,
the open circles for intermediate ellipticals,
the open squares for bright dwarf ellipticals,
the filled triangles for compact ellipticals,
and the open pentagons for dwarf spheroidals.
The solid lines are the best-fit relations for the
giant-dwarf elliptical sequence (GDE) and 
spiral galaxies, while the dot-dashed line is the best-fit to the
giant-compact elliptical sequence (GCE). The dashed lines are the
relation shifted by standard deviation in $\Delta(\log r_e)$
from the solid lines. 
}
\label{fig:L-Re}
\end{figure}

In our PLE model, we assume that the galaxy size does not evolve 
except for the case of mergers of galaxies,
and use the empirical relation between the effective radius $r_e$ and 
absolute luminosity $L_B$ for local galaxies. 
If we allow for the number evolution of galaxies, we must take into 
account the change of galaxy sizes, and we will discuss this in 
\S \ref{sect:merger}.  
Fig. \ref{fig:L-Re} shows the size-luminosity relation of local 
elliptical and spiral galaxies. 
The data of elliptical galaxies are taken from
Bender et al. (1992), and those of spiral galaxies
from Impey et al. (1996). 
Although there is a significant scatter in this relation,
we use a simple power-law relation for this relation as
\begin{equation}
r_e \propto L_B^{2.5/p} \ ,
\label{eq:LB-re}
\end{equation}
or, if expressed in terms of the absolute $B$ magnitude,  
\begin{equation}
- M_B = p \log r_e + q + (p-5) \log (H_0/50\ \rm km/s/Mpc) \ .
\end{equation}
Elliptical galaxies form two distinct families which 
follow the well-separated sequences at low luminosities 
in the $r_e$-$L_B$ diagram.
One is the ordinary sequence from giant through dwarf elliptical 
(GDE) galaxies, while the other is the bright sequence from giant 
through compact elliptical (GCE) galaxies (see
Fig \ref{fig:L-Re}). Since the predictions of galaxy number count
are not sensitive to whichever sequence is used in the analysis
(see \S \ref{section:model-dependence} and 
Fig \ref{fig:nm-r-e}), we take the GDE sequence as the 
standard size-luminosity relation of elliptical galaxies in this 
paper.  

The $r_e$-$L_B$ relations fitted to the data in Fig.\ref{fig:L-Re}  
yield $(p,q)=(6.0,16)$ and (3.5,18.7) for the GDE and GCE sequences
of elliptical galaxies, respectively, and (9.4, 12) for spiral 
galaxies.  In order to examine the uncertainty due to the significant 
scatter in the $r_e$-$L_B$ relation, we derive the standard deviation 
in $\Delta(\log r_e)$ from the best-fit relation and repeat 
calculations with the shifted relations shown in Fig. \ref{fig:L-Re} 
by the dashed lines.

The radial distribution of surface brightness is assumed to follow 
de Vaucouleurs' (1962) profile ($S \propto \exp[-(r/r_e)^{1/4}$] ) for 
elliptical galaxies and an exponential profile for spiral galaxies 
(Freeman 1970).  Then we can calculate the radial distribution of 
surface brightness of a galaxy at given redshift in any passband, when 
the galaxy type, the present-day $B$ luminosity, and the evolution model 
are specified. This surface brightness profile should be convolved with 
a Gaussian point-spread function (PSF) having the same dispersion with 
the observational seeing.

\subsubsection{Object Detection}

Let $S_{\rm th}$ be the surface brightness threshold adopted in a galaxy 
survey. Strictly, 
this threshold could change due to different noise levels within a survey
field, but we use a single value for the simplicity.  When the observed 
profile of a galaxy image is calculated as above, we can estimate the 
isophotal size which encircles a region of a galaxy image with surface 
brightness brighter than $S_{\rm th}$.  If the isophotal size of a galaxy 
is zero, i.e., the central surface brightness is fainter than $S_{\rm th}$, 
this galaxy can not be detected in the survey.  Usually a minimum isophotal 
diameter $D_{\rm min}$, which is comparable to the seeing size, is adopted
as a condition for an image to be detected as a galaxy.  We can calculate 
the isophotal diameter for a model galaxy, then it is easy to check whether 
this galaxy meets the detection criterion in the galaxy survey.

\subsubsection{Photometry Scheme}
There are three photometry schemes to evaluate the apparent magnitude of 
galaxies such as isophotal, aperture, and pseudo-total magnitudes. These 
magnitudes may significantly differ especially near the detection limit, 
and this difference should be included in the theoretical modeling of the
selection effects inherent in the method of detecting faint galaxies.

The isophotal magnitude is the flux within the isophotal size of a galaxy 
image.  The aperture magnitude is the flux within a fixed aperture
which is adopted by observers.  Some observers often make corrections
to these magnitudes into pseudo-total magnitudes, which are intended
to mimic an ideal total flux of a galaxy.  However, such corrections use 
a model luminosity profile to evaluate the flux from an `undetected' 
part of a galaxy image.  It is inappropriate to make a model-dependent
correction to the observed 
quantities with which various theoretical models are compared.  The best 
way is, on the contrary, to incorporate all the necessary corrections 
in the theoretical models to be compared directly with the observed
quantities.  We therefore suggest that observers should also present 
raw counts and magnitudes, in addition to presenting corrected quantities
in their papers.

\subsubsection{Detection and Photometry of the HDF Galaxies}
\label{sect:HDF-selection}
Here we describe the detection processes of the HDF galaxies, following 
Williams et al. (1996), and see also Bouwens, Broadhurst, \& Silk (1998).  
Object detection is performed in the combined 
$V_{606}+I_{814}$ image. It is first convolved with a fixed smoothing 
kernel of 25 pixels (=0.04 square arcseconds), then pixels having values 
higher than a fixed threshold above a local sky background are marked 
as potentially being part of an object.  After thresholding, regions 
consisting of more than contiguous 25 pixels are counted as sources. 
It corresponds to an isophotal diameter limit of $D_{\min} \sim$ 0.2 
arcsec, about 1.6 times larger than the FWHM of the PSF. 

\begin{figure}
  \begin{center}
    \leavevmode\psfig{file=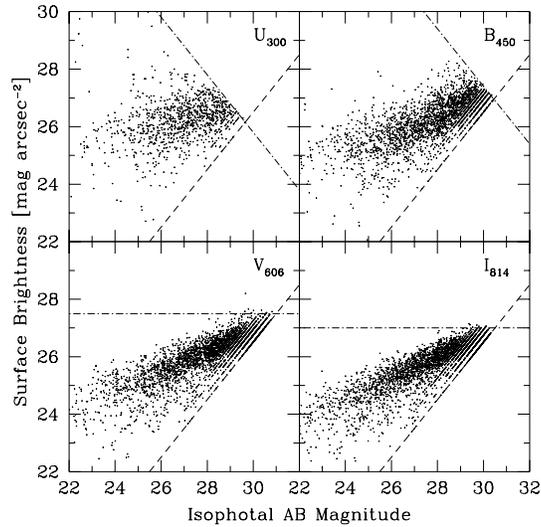,width=7.3cm}
  \end{center}
\caption{Isophotal AB magnitude versus average surface brightness of
the HDF galaxies (Williams et al. 1996). The dashed line is the limit of
minimum area of a galaxy image $A = 0.04 \ \rm arcsec^{-2}$. 
The dot-dashed lines
in the $V_{606}$ and $I_{814}$ bands are the adopted isophotal 
thresholds ($S_{\rm th}$ = 27.5 and 27.0 mag arcsec$^{-2}$ for
the $V_{606}$ and $I_{814}$ bands, respectively). The dot-dashed lines
in the $U_{300}$ and $B_{450}$ bands correspond to the detection 
threshold by signal to noise ratio of $S/N>2$ (see text).
}
\label{fig:mag-S}
\end{figure}

The surface brightness threshold is not clearly indicated in Williams et 
al. (1996), but we can evaluate this value from the surface brightness 
distribution of the HDF galaxies.  In Fig. \ref{fig:mag-S}, we plot 
their apparent magnitudes versus average surface brightness in 
the four passbands of $U_{300}$, $B_{450}$, $V_{606}$, and $I_{814}$ 
by using the published sizes and magnitudes of the HDF catalog.  Here 
the size refers to the isophotal size of the combined $V_{606}+I_{814}$ 
image, and the magnitude refers to the isophotal magnitude measured within the 
isophotal size. First we consider the $V_{606}$ and $I_{814}$ bands.
Figure \ref{fig:mag-S} shows that no galaxies are
detected when the 
average surface brightness is fainter than $S_{\rm th}= 27.5$ mag arcsec$^{-2}$
in $V_{606}$ and $27.0$ mag arcsec$^{-2}$ in $I_{814}$.
The faintest surface brightness detected in the HDF for galaxies
with a fixed isophotal magnitude becomes 
brighter for brighter galaxies.  This trend occurs because the size of 
galaxies is larger for brighter galaxies.  The edge of an isophotal image 
corresponds to the isophotal limit, and hence its average surface 
brightness within the isophotal area is always brighter than its threshold
when the galaxy image have a bright central part and therefore a larger
size.  Consequently the faintest surface brightness at the faintest isophotal 
magnitudes, i.e., $S_{\rm th}= 27.5$ mag arcsec$^{-2}$
in $V_{606}$ and $27.0$ mag arcsec$^{-2}$ in $I_{814}$,
gives the surface brightness threshold for object detection.
In our calculations the above threshold values in the $V_{606}$ and $I_{814}$ 
bands are used respectively, although in reality the detection was done by 
the combined $V_{606}+I_{814}$ image in the HDF catalog.

On the other hand, the surface brightness threshold is not clear in
the $U_{300}$ and $B_{450}$ bands, because the object detection was done 
without these bands.  It should also be noted that the isophotal $U_{300}$ 
and $B_{450}$ magnitudes are defined as the flux within the isophotal size 
in the combined $V_{606}+I_{814}$ image.  Among the objects to which these
isophotal magnitudes are assigned, those with $S/N>2$ are detected as galaxy
images in the $U_{300}$ and $B_{450}$ bands (Williams et al. 1996).  We see 
a clear boundary indicated by dot-dashed line running from upper-left to 
lower-right in the $U_{300}$ and $B_{450}$ panels of Fig. \ref{fig:mag-S}, 
and this corresponds to the line of $S/N=2$.  It is easy to show that this 
is equivalent to a condition of $S+m$ = const if the noise level is 
proportional to $A^{1/2}$ (Poisson type noise), where $A$ is the isophotal 
area.  We have also confirmed that the 
galaxies in $U_{300}$ and $B_{450}$ with $S/N \sim 2$
are actually on the dot-dashed line of Fig. \ref{fig:mag-S}
\footnote{This is true for the three WF fields, but not for the PC field, 
because of the different surface brightness threshold employed (Williams et 
al. 1996).  The number of galaxies in the PC field is negligible compared 
with the WF fields.}.
In our calculations the galaxies in the $U_{300}$ and $B_{450}$ bands are
detected if they are detected in $I_{814}$ band and furthermore meet the 
criterion of $S/N>2$ within the isophotal area in the $I_{814}$ band (i.e., 
those below the dot-dashed line in Fig. \ref{fig:mag-S}).

\subsection{Merger and Number Evolution}
\label{sect:merger}
Currently the most popular theory for the structure formation in the 
universe is the bottom-up scenario with the cold dark matter which 
dominates the total mass density of the universe, in which smaller mass 
objects form earlier and then merge into larger objects 
(e.g., Blumenthal et al. 1984).
Merging history of dark matter haloes is relatively 
well studied by analytical methods as well as $N$-body simulations, but
merging history of galaxies could significantly differ from that of dark 
matter haloes and is poorly known.  

Since a number evolution of galaxies caused by galaxy mergers significantly 
affects the number count of faint galaxies, we investigate this effect 
by using a simple merging model in which the luminosity density of galaxies 
is conserved.  A common practice for this is to introduce the 
redshift-dependent parameters of Schechter-type luminosity function of 
galaxies such as 
\begin{eqnarray}
\phi^*(z) &=& \phi^*(0) (1+z)^\eta  \nonumber \\	
L^*(z) &=& L^*(0) (1+z)^{-\eta} \ .
\label{eq:schechter}
\end{eqnarray}
We adopt a single value of $\eta$ for all types of galaxies for simplicity. 

In the analysis of this paper, the size of a galaxy is crucially important
for evaluating the selection effects.  Merger of galaxies should change 
their size, and this should also be taken into account.  The empirical 
relation $r_e \propto L_B^{2.5/p}$ may not hold at high redshifts, 
depending on how $r_e$ and $L_B$ change during the merger process. 
We assume that the change of
$L_B$ and $r_e$ during merger processes always satisfy a relation $L_B \propto 
r_e^\xi$. (Note that this relation is physically
different from equation \ref{eq:LB-re} which is the relation of galaxies at a
fixed time, but describing the change of luminosity
and size of a test galaxy during merger processes.)
If $\xi=2$, the surface brightness of galaxies is conserved 
during mergers, and if $\xi=3$, the luminosity density in each galaxy is 
conserved.  Then the luminosity and size of a $z=0$ galaxy evolve as 
\begin{eqnarray}
L_B &\rightarrow& L_B (1+z)^{-\eta} \\
r_e &\rightarrow& r_e (1+z)^{-\eta/\xi} \ .
\end{eqnarray}
By applying this transformation to the empirical $r_e$-$L_B$ relation at 
$z=0$, it is straightforward to give the $r_e$-$L_B$ relation as
a function of redshift:
\begin{equation}
r_e(L_B, z) = r_e(L_B, 0) \times (1+z)^{-\frac{\eta}{p}
\left(\frac{p}{\xi} - 2.5\right)} \ ,
\end{equation}
which is a generalization of eq. \ref{eq:LB-re}.

The value of $\xi$ depends on the physical process of mergers.  Generally, 
merger products are expected to become more compact than pre-merger 
progenitors in gas-rich mergers because of efficient cooling and 
dissipation.  On the other hand, merger products become less compact 
in gas-less mergers because the relative translational energy of pre-merger
stellar progenitors is converted into the internal kinetic energy of a merged 
stellar system.  The former corresponds to larger $\xi$, while the latter to 
smaller $\xi$.  We use $\xi=3$ as a standard value and examine the effect of 
changing this value.  The assumed conservation of the total luminosity density 
of all galaxies (i.e., $\phi^*(z) L^*(z)$ = constant) 
will also be discussed when we draw our conclusion of this paper.

\subsection{Formulations}
In the following we describe the formulations necessary to calculate the 
number count and redshift distribution of faint galaxies.  We denote
$\lambda$ as the observed wavelength, and $\lambda_D$ as the wavelength
at which the object detection is performed.  For example, we use 
$\lambda=\lambda_D=V_{606}$ or $I_{814}$ for counts in the
$V_{606}$ and $I_{814}$ bands, whereas $\lambda_D=I_{814}$ 
and $\lambda=U_{300}$ or $B_{450}$ for the $U_{300}$ and $B_{450}$ bands
(see \S \ref{sect:HDF-selection}).

The number of galaxies per unit steradian, per unit apparent magnitude
($m_\lambda$), and per unit redshift is written as
\begin{equation}
\frac{d^3N}{dm_\lambda dzd\Omega} = H(x)
\frac{d^2V}{dzd\Omega} \sum_i \phi_i(M_B, z)
\frac{dM_B}{dm_\lambda} \ , 
\label{eq:dNdmdz}
\end{equation}
where $x=D_{\lambda_D}-D_{\min}$, $D_{\lambda_D}$ is the isophotal 
diameter of galaxies measured in the $\lambda_D$ band, $d^2V/dzd\Omega$ 
is the comoving volume element which depends on the cosmological 
parameters, $\phi_i$ is the luminosity function per unit $M_B$ with the 
Schechter parameters given in equation (\ref{eq:schechter}), and $H$ is 
the step function [$H(x) = 1$ and 0 for $x \geq 0$ and $<0$, 
respectively].  The absolute $B$ magnitude $M_B$ is that of the present-day
galaxies, which is related with $z$ and $m_\lambda$ by $K$-correction
and evolutionary $(E)$ correction (see equation \ref{eq:m-lambda}
below).  The subscript $i$ 
denotes the galaxy type.  The quantity $d^3N/dm_\lambda dz d\Omega$ gives 
the redshift distribution of galaxies, and the integration over $z$ gives 
the number count or the number-magnitude relation.

We calculate $D_{\lambda_D}$ and $M_B$ as a function of $m_\lambda$ and 
$z$, taking into account the selection effects and the photometry scheme.  
This is carried out as follows:  For a given set of $M_B$ and $z$, by 
using the assumed luminosity profile and the $L_B$-$r_e$ relation, we 
first calculate the surface brightness distribution of a galaxy image in 
the $\lambda_D$ band. Comparing this surface brightness distribution with 
the adopted threshold $S_{\rm th}$ 
in the $\lambda_D$ band, we then calculate the 
isophotal diameter $D_{\lambda_D}$.  Given this diameter, we calculate the 
isophotal magnitude in the $\lambda$ band. 
(If the aperture magnitude is used in a survey, 
we should use the fixed aperture here.  The pseudo-total magnitude is 
easily calculated simply from the total absolute magnitude without using 
the surface brightness distribution.)  In this way we finally obtain 
$m_\lambda$ as a function of $M_B$ and $z$, or conversely $M_B$ can 
be related to $m_\lambda$.  It is obvious that $D_{\lambda_D}$ is 
automatically obtained in this process.  In the following we give
detailed numerical formulations necessary for the above calculations.

Let $g(\beta)$ be radially symmetric luminosity profile (surface brightness 
distribution) of a galaxy, where $\beta = r/r_e$ is the radius from the 
center normalized by the effective radius of the galaxy.  The adopted form
of the profile is given by   
\begin{equation}
g(\beta) = \exp( - a_n \beta^{1/n}) \ ,
\end{equation}
and the integrated profile out to $\beta$ is given by  
$G(\beta) \equiv 2 \pi \int_0^\beta g(\beta')\beta' d\beta' \ .$
We assume de Vaucouleurs' profile ($n=4$) for elliptical galaxies and
an exponential profile ($n=1$) for spiral galaxies, as mentioned
earlier.  An effective radius $r_e$ is defined as the radius within
which a half of total luminosity is encircled, and by this definition 
the coefficient $a_n$ is given by $a_4 = 7.67$ and $a_1 = 1.68$.
In order to incorporate the effect of observational seeing, we 
convolve this profile function with a Gaussian PSF with dispersion 
$\sigma_t$:
\begin{eqnarray}
\tilde{g}(\beta) &=& \int_0^\infty \frac{d\xi \xi}{\sigma_t^2}
g(\xi)\left\{ I_0\left(\frac{\beta\xi}{\sigma_t^2}\right)
\exp\left(-\frac{\beta\xi}{\sigma_t^2}\right)\right\} \nonumber \\
&\times&
\exp\left(-\frac{(\beta-\xi)^2}{2\sigma_t^2}\right) \ ,
\end{eqnarray}
where $I_0(x)$ is the modified Bessel function of the first kind
(e.g., Press et al. 1992).  The seeing FWHM in units of radian is 
now related to $\sigma_t$ as
\begin{eqnarray}
\sigma_t = \left(\frac{\rm seeing \ FWHM}{2.35}\right) 
\frac{d_A(z)}{r_e} \ ,
\end{eqnarray}
where $d_A(z)$ is the standard angular diameter distance.

The surface brightness distribution of a galaxy image is given by
\begin{eqnarray}
S_\lambda(\theta)[{\rm mag \ arcsec^{-2}}] &=& M_B + (\lambda-B)_0
+ K_\lambda(z) + E_\lambda(z) \nonumber \\ 
&+& 2.5 \tau_{\rm HI}(\lambda, z) \log e \nonumber \\
&+& 5 \log [r_e(M_B, z)(1+z)^2/{\rm 10 pc}] \nonumber \\
&+& 26.5721 \nonumber \\
&-& 2.5 \log [\tilde{g}(\beta)
/ G(\infty)] \ ,
\end{eqnarray}
where $\theta$ is the angular radius from the center of a galaxy image,
$K_\lambda$ is the $K$-correction, $E_\lambda$ is the $E$-correction 
including internal absorption by dust, and $\tau_{\rm HI}$ is the optical 
depth of intergalactic absorption by HI clouds. 
The present-day color of galaxies of a given type,
$(\lambda-B)_0$, is calculated by the AB colors based on the
present-day SEDs of model galaxies
which reproduce the observed colors, and we use $B_{450} - B = -0.153$
to relate the AB magnitudes to the $B$ magnitudes of the
luminosity function. The radius parameter $\beta$ 
is related to $\theta$ as $\beta = d_A\theta / r_e$, and 
$G(\infty) = (2n)! \pi / a_n^{2n}$.  The isophotal size $D_{\lambda_D}$ 
can be derived from solving the equation $S_{\lambda_D}(D_{\lambda_D}/2) = 
S_{\rm th}$.  Similarly, the apparent magnitude encircled within $\theta$
is given by
\begin{eqnarray}
m_\lambda(\theta) &=& M_B + (\lambda-B)_0
+ K_\lambda(\lambda, z) + E_\lambda(\lambda, z)  \nonumber \\
&+& 2.5 \tau_{\rm HI}(\lambda, z) \log e
+ 5 \log [d_L/{\rm 10 pc}] \nonumber \\
&-& 2.5 \log [\tilde{G}(\beta)
/ G(\infty)] \ ,
\label{eq:m-lambda}
\end{eqnarray}
where $d_L$ is the standard luminosity distance, and $\tilde{G}(\beta)$
is the integrated profile of $\tilde{g}(\beta)$.  The isophotal, aperture, 
and pseudo-total magnitudes are obtained with $\theta = D_{\lambda_D}/2$, 
aperture/2, and $\infty$, respectively. (Note that $\tilde{G}(\infty)
= G(\infty)$.)

The $K$ and $E$ corrections are calculated from the time-dependent models 
of spectral energy distribution per unit wavelength $f_\lambda(t)$, 
after modified to include the dust absorption.  They can be written as
\begin{eqnarray}
K_\lambda(z) &=& -2.5 \log \left\{ \frac{1}{(1+z)}\frac{f_{\lambda/(1+z)}
(t_0)}{f_\lambda (t_0)} \right\} \ , \\
E_\lambda(z) &=& -2.5 \log \left\{ \frac{f_{\lambda/(1+z)}(t_z)}
{f_{\lambda/(1+z)}(t_0)} \right\} \ ,
\end{eqnarray} 
where $t_z$ is the age of a galaxy at redshift $z$ which was formed
at $z_F$, and $t_0$ is its present age.

With all these prescriptions, we can numerically solve $M_B$ for a 
given set of $m_\lambda$ and $z$, and then the number count and redshift 
distribution of galaxies by means of equation (\ref{eq:dNdmdz}).

\section{Faint Galaxy Number Counts}
\label{section:counts}
\subsection{Importance of the Observational Selection Effects}
Figure \ref{fig:nm-type} shows the predictions of galaxy number count
in the HDF based on the PLE model of galaxy evolution, including the 
effects of dust and intergalactic absorptions, and the selection effects.  
Here we have used a `standard' set of model parameters in this paper: 
$(h, \Omega_0, \Omega_\Lambda)=(0.7, 0.2, 0.8)$, $z_F=5$, the local 
luminosity function of the SSRS2 survey, and the screen model of 
dust. [Here, $h = H_0$/(100km/s/Mpc) as usual.]
The solid line is the total number count of all galaxy types,
and the other five lines are those of respective galaxy types.
In this paper we will present many calculations of galaxy counts,
changing various parameters in order to check the systematic uncertainties
which may affect our conclusions. The summary of count calculations
is presented in Table \ref{table:nm-summary} with references to the number 
of figures in this paper.

In the following we compare these calculated counts with the HDF data as 
well as the ground-based data transformed into the AB magnitude system.  
Here we note that the HDF bandpass filters for the $U_{300}$ and $V_{606}$ 
are significantly different from those for the $U$, $V$, and $R$ used 
in the ground-based observations.  We have corrected the magnitudes of 
such ground-based data, making use of the central wavelength of bandpass 
filters, the present-day SEDs of different galaxy types, and the relative 
proportions among the galaxy types. The HDF counts are those of
isophotal magnitudes, while published total-magnitude counts are used for
other ground-based data for which the selection effects are not as
important as for the HDF data.

Figure \ref{fig:nm-sel-nosel} shows the effect of cosmological parameters
and the selection effects in the predictions of galaxy number count.  
The predictions with the selection effects are presented by solid lines, 
while those without the selection effects by dashed lines.  In either 
cases, the three lines from bottom to top correspond to the EdS universe 
with $(h, \Omega_0, \Omega_\Lambda)=(0.5, 1, 0)$, an open universe with 
$(h, \Omega_0, \Omega_\Lambda)=(0.6, 0.2, 0)$, and a $\Lambda$-dominated 
flat universe with $(h, \Omega_0, \Omega_\Lambda)=(0.7, 0.2, 0.8)$. 

The dotted line is a prediction without the selection effects in an 
open universe with $\Omega_0=0.1$.  The prescriptions used here are the 
same as those in the prediction by Pozzetti et al. (1998) which they 
found to agree with the observed HDF counts.  Our prediction shown by 
this dotted line also gives a good fit to the HDF data, and we have 
confirmed Pozzetti et al.'s result where no selection effects are taken 
into account.  This suggests that the difference between the PLE models 
of different authors is not significant in the number count predictions. 

However, this figure clearly demonstrates the importance of the 
selection effects in comparison between the predicted and observed 
number counts of galaxies.  The difference between the predicted counts 
with and without the selection effects attains up to a factor of about 
4 at the faintest magnitudes.  As a result, the observed HDF counts in
excess of the PLE predictions with the selection effects are much
larger than previously considered.  It is striking that such PLE 
predictions seem to be short of the observed counts in all the 
passbands, even in the $\Lambda$-dominated flat universe with 
$\Omega_\Lambda$ = 0.8 in which the number count is close to the 
maximum.  In the EdS universe, such a predicted deficit in the PLE 
model attains up to a factor of more than 10.  This large excessive
number of HDF galaxies may suggest the number evolution of galaxies, 
but it depends heavily on the cosmology
how much the number evolution is required to explain the HDF data.

\subsection{Dependence on Model Parameters}
\label{section:model-dependence}
\begin{figure}
  \begin{center}
    \leavevmode\psfig{file=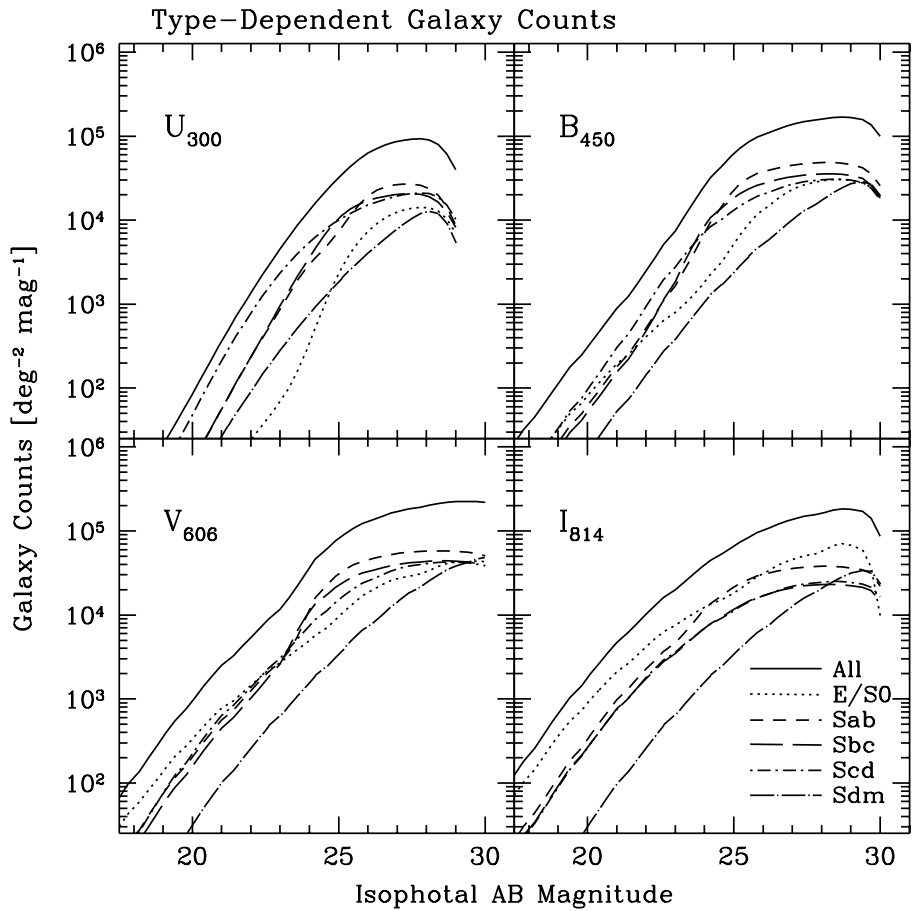,width=13cm}
  \end{center}
\caption{Faint galaxy number counts predicted in the four bands
of the HDF. The selection effects under the HDF observation condition 
are included. The model presented here is our `standard' PLE model with
the standard set of model parameters: $(h, \Omega_0, \Omega_\Lambda) = 
(0.7, 0.2, 0.8)$, $z_F = 5$, the local luminosity function of the SSRS2 
survey, and the screen model of dust extinction. Intergalactic absorption 
by HI clouds is also taken into account. 
The solid line is the total counts of all five types of galaxies,
while the other lines 
are for individual types of E/S0 (dotted), Sab (short-dashed), 
Sbc (long-dashed), Scd (dot-short-dashed),
and Sdm (dot-long-dashed). 
}
\label{fig:nm-type}
\end{figure}

\begin{figure}
  \begin{center}
    \leavevmode\psfig{file=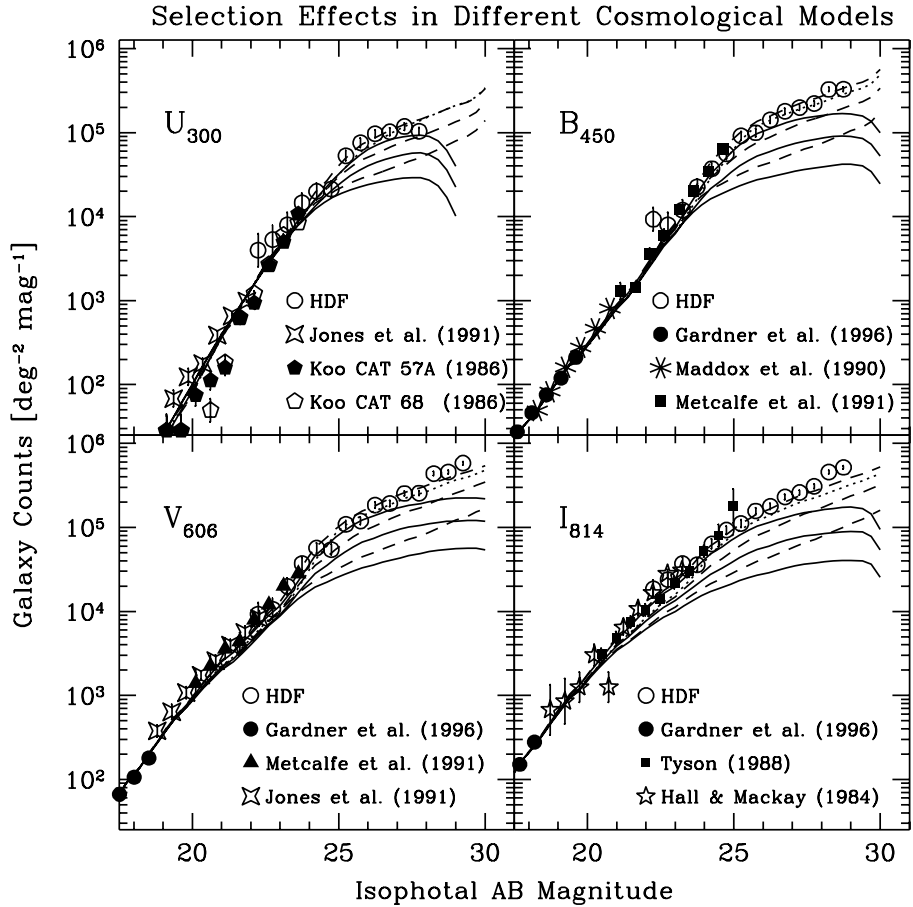,width=13cm}
  \end{center}
\caption{Comparison of the PLE model with the observed HDF counts as
well as ground-based brighter counts. The top solid line is the 
standard PLE model shown in Fig. \ref{fig:nm-type} in a $\Lambda$-dominated
flat universe with
$(h, \Omega_0, \Omega_\Lambda) = (0.7, 0.2, 0.8)$, while the middle
and bottom lines are for the different cosmologies of an open universe
with (0.6, 0.2, 0.0) and an EdS universe with (0.5, 1, 0),
respectively. The three dashed lines from top to bottom
are the same as the solid lines,
except that the observational selection effects are not taken into account.
The dotted line is a prediction in an open universe with
$(h, \Omega_0, \Omega_\Lambda) = (0.6, 0.1, 0.0)$ and no selection
effects, which are the same prescription with 
the previous study by Pozzetti et al. (1998).
}
\label{fig:nm-sel-nosel}
\end{figure}

\begin{figure}
  \begin{center}
    \leavevmode\psfig{file=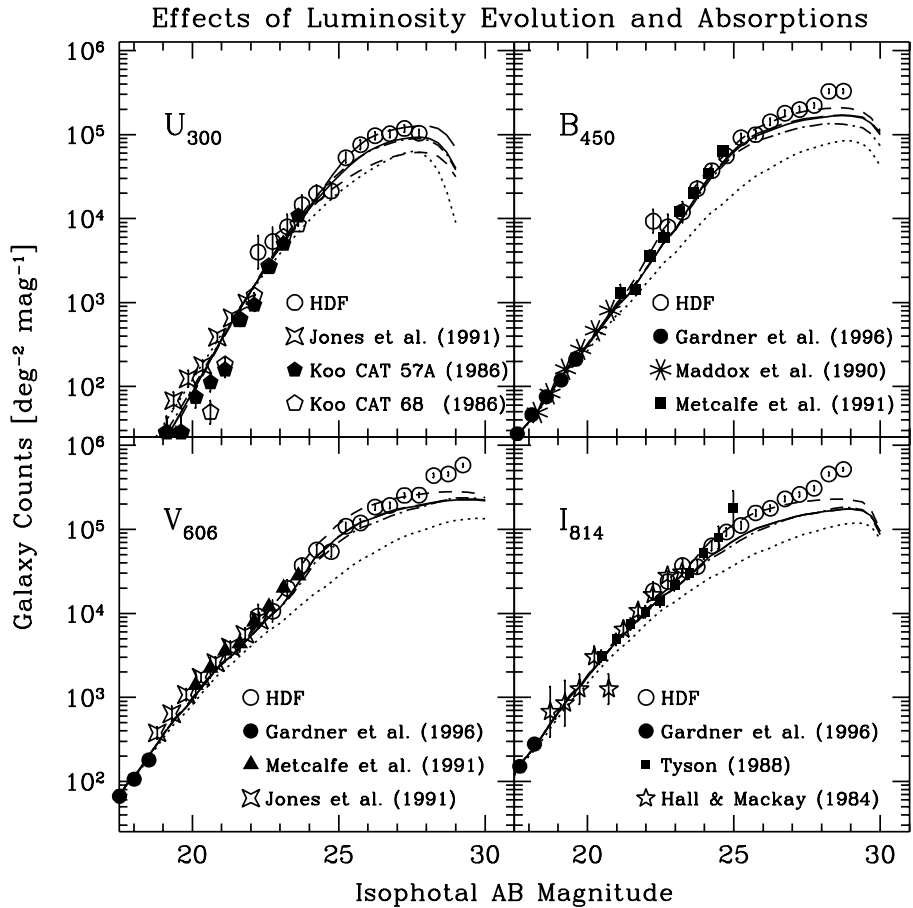,width=13cm}
  \end{center}
\caption{Comparison of the PLE model with the observed HDF counts as
well as ground-based brighter counts. The solid line is the 
standard PLE model shown in Fig. \ref{fig:nm-type}. The other lines
are the same but for the case of no luminosity evolution (dotted),
slab model dust (short-dashed), no intergalactic absorption by HI clouds
(long-dashed). The dot-dashed line is for the PLE model with the
updated luminosity evolution model of Kobayashi et al. (1999).
}
\label{fig:nm-model}
\end{figure}

\begin{figure}
  \begin{center}
    \leavevmode\psfig{file=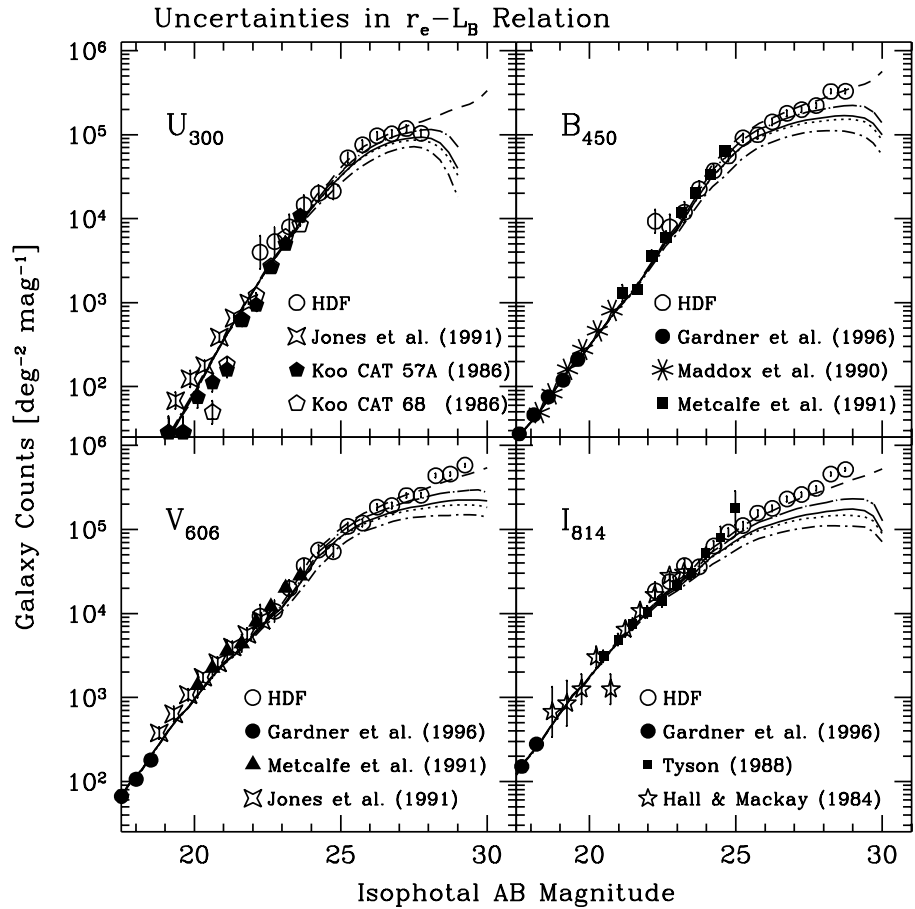,width=13cm}
  \end{center}
\caption{Comparison of the PLE model with the observed HDF counts 
as well as ground-based brighter counts. The solid line is the standard 
PLE model shown in Fig. \ref{fig:nm-type}.  The other lines are 
the same but for models without the selection effects (dashed)
and with the shifted size-luminosity relation by $+1\sigma$ 
(short-dot-dashed) and $-1\sigma$ (long-dot-dashed) deviation 
in $\Delta (\log r_e)$.
The dotted line is the PLE model where the GCE sequence of elliptical
galaxies is used instead of the standard GDE sequence.
See \S \ref{section:size-luminosity} for detail.
}
\label{fig:nm-r-e}
\end{figure}

Before we investigate the effect of number evolution quantitatively, 
it is necessary to see the uncertainties in the PLE predictions.
Figure \ref{fig:nm-model} shows how the PLE predictions depend on 
models of luminosity evolution and absorptions.  The solid line is the 
standard model presented in Fig. \ref{fig:nm-type}.  The dotted line 
is the same as the standard model but with no luminosity evolution.  
The dot-dashed line is the prediction where the updated luminosity 
evolution model of Kobayashi et al. (1999) is used rather than the 
standard model of Arimoto \& Yoshii (1987) and Arimoto, Yoshii, \& 
Takahara (1992).  The short dashed line is the prediction where the 
slab distribution of dust is assumed rather than the screen by dust.  
The long-dashed line is the prediction where no intergalactic 
absorption is taken into account.  This line almost overlaps with the 
solid line and is not visible except in the $U_{300}$ band.  

\begin{figure}
  \begin{center}
    \leavevmode\psfig{file=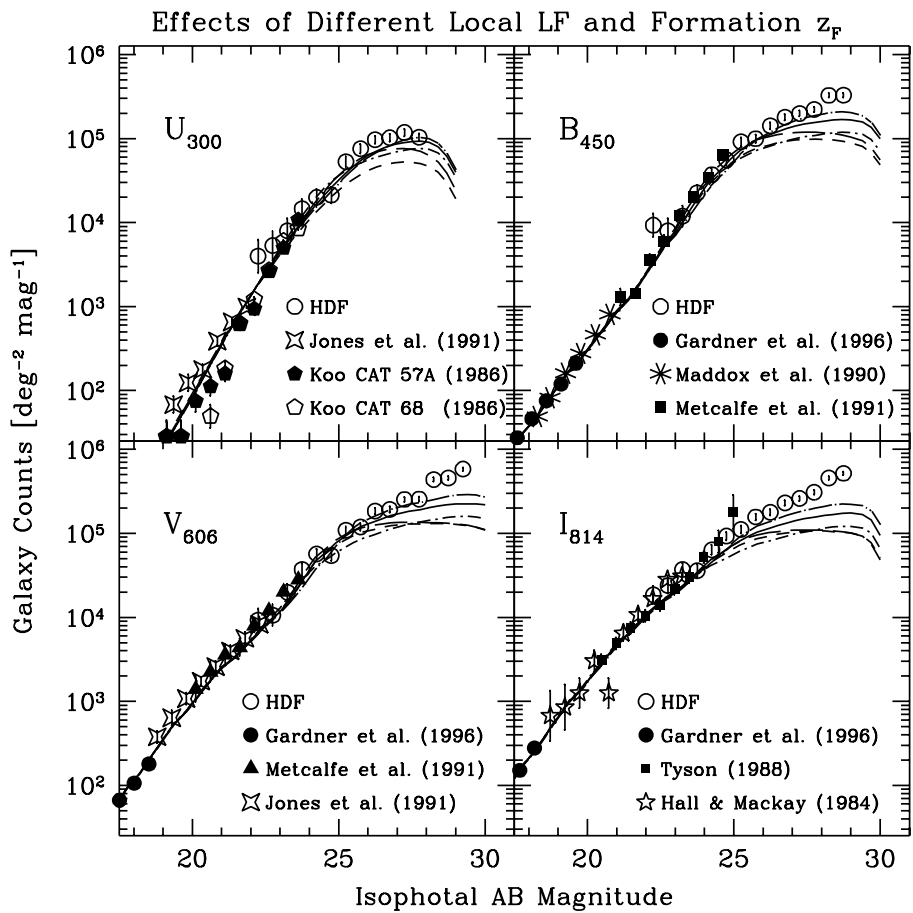,width=13cm}
  \end{center}
\caption{Comparison of the PLE model with the observed HDF counts as
well as ground-based brighter counts. The solid line is the 
standard PLE model shown in Fig. \ref{fig:nm-type}. The other lines
are the same but for the models with $z_F = 3$ (short-dot-dashed) and 
10 (long-dot-dashed), and with different local luminosity functions of
the Stromlo-APM survey (short-dashed) and the CfA survey (long-dashed).
}
\label{fig:nm-zFlf}
\end{figure}

Figure \ref{fig:nm-zFlf} shows the effect of changing the galaxy 
formation epoch $z_F$ and local luminosity function.  The solid line 
is the standard model presented in Fig. \ref{fig:nm-type}.  The
short- and long-dot-dashed lines are the predictions with $z_F$ = 3 
and 10, respectively, rather than $z_F=5$ in the standard model. 
The short- and long-dashed lines are the predictions where the 
luminosity function of the Stromlo-APM and the CfA redshift 
surveys are used, respectively, with $z_F=5$. 

Figure \ref{fig:nm-r-e} shows how the uncertainty in the
size-luminosity relation affects the predictions of galaxy number
count. The solid line is the standard model presented in 
Fig. \ref{fig:nm-type}, while the dashed line is the prediction
when the selection effects are ignored,
as shown in Fig. \ref{fig:nm-sel-nosel}. 
The short- and long-dot-dashed
lines are the counts when the $r_e$--$L_B$ relation is shifted
by $\pm$ 1 $\sigma$ in $\Delta(\log r_e)$. The dotted line is 
the prediction using the GCE sequence instead of the standard GDE
sequence as the size-luminosity relation of elliptical galaxies.
(See \S \ref{section:size-luminosity} for detail.)

From these three figures we understand a range of the systematic
uncertainties in the PLE predictions, which turns out to be not 
large enough to save the EdS universe nor an open universe with 
$\Omega_0>0.2$.  These uncertainties are significant at the faintest 
magnitudes where the $N$-$m$ relation starts to turn over.  However, 
we point out that the effect of cosmological parameters becomes 
apparent at brighter magnitudes where the uncertainties remain much 
less significant.  In fact, we will show that the slope of the 
$N$-$m$ relation at $B_{450}$ = 22--26 can be used to discriminate 
between the effects of cosmological parameters and galaxy number 
evolution. In order to demonstrate the above statement quantitatively,
the systematic model uncertainties at $m$= 25 and 28 are summarized in
Table \ref{table:nm-sensitivity}. The uncertainty at $m=25$, to be 
compared with the effect of cosmological parameters, is dominated by 
the dust distribution model, but we note that the estimated change 
from a standard screen model to the slab model is somewhat overestimated 
because the slab model is clearly inconsistent as a model of dust
distribution in starburst galaxies, as mentioned in \S 
\ref{section:absorption}.

\subsection{Mergers and Number Evolution}
\begin{figure}
  \begin{center}
    \leavevmode\psfig{file=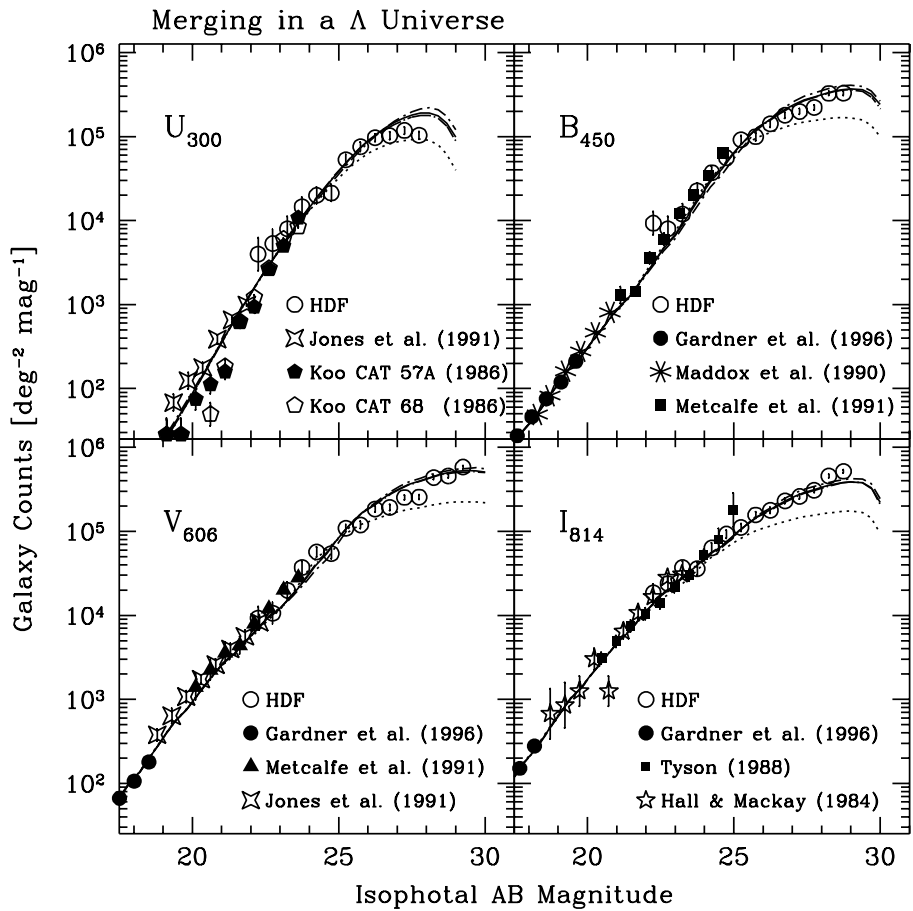,width=13cm}
  \end{center}
\caption{The effect of galaxy number evolution on galaxy counts based 
on the luminosity-density conserving mergers in a $\Lambda$-dominated
universe with $(h, \Omega_0, \Omega_\Lambda) = (0.7, 0.2, 0.8)$. 
The dotted line is the standard PLE model without number evolution
in this universe shown in Fig. \ref{fig:nm-type} as well as Fig.
\ref{fig:nm-sel-nosel}. The short-dot-dashed, solid, and long-dot-dashed 
lines are the models with the merger parameters of $(\eta, \xi)=$(1, 2), 
(1, 3), and (1, 4), respectively.
}
\label{fig:nm-lam-merge}
\end{figure}

The PLE predictions fall considerably short of the observed HDF 
counts in the EdS universe and an open universe, and it is still the 
case even in a $\Lambda$-dominated flat universe.  Here, by using a 
simple model of mergers introduced in \S \ref{sect:merger}, 
we investigate whether
the number evolution explains the large number of faint HDF galaxies. 
Figure \ref{fig:nm-lam-merge} shows the effect of introducing such a
number evolution model in a $\Lambda$-dominated flat universe.
The solid line is the prediction with the merger parameters of 
$(\eta, \xi) = (1, 3)$, while the dotted line is the standard PLE 
prediction with no number evolution (Fig. \ref{fig:nm-type}). 
The short- and long-dot-dashed lines are the predictions with 
$(\eta, \xi) = (1, 2)$ and (1, 4), respectively, showing that the 
effect of changing $\xi$ is not significant.  
This result indicates that a modest number evolution with $\eta 
\sim 1$ [$\phi^* \propto (1+z)^{\eta}$] is sufficient to explain the 
observed HDF counts in a $\Lambda$-dominated flat universe with
$\Omega_\Lambda=0.8$, and an even stronger number evolution with 
$\eta \gtilde 1$ is rejected by the data. 

\begin{figure}
  \begin{center}
    \leavevmode\psfig{file=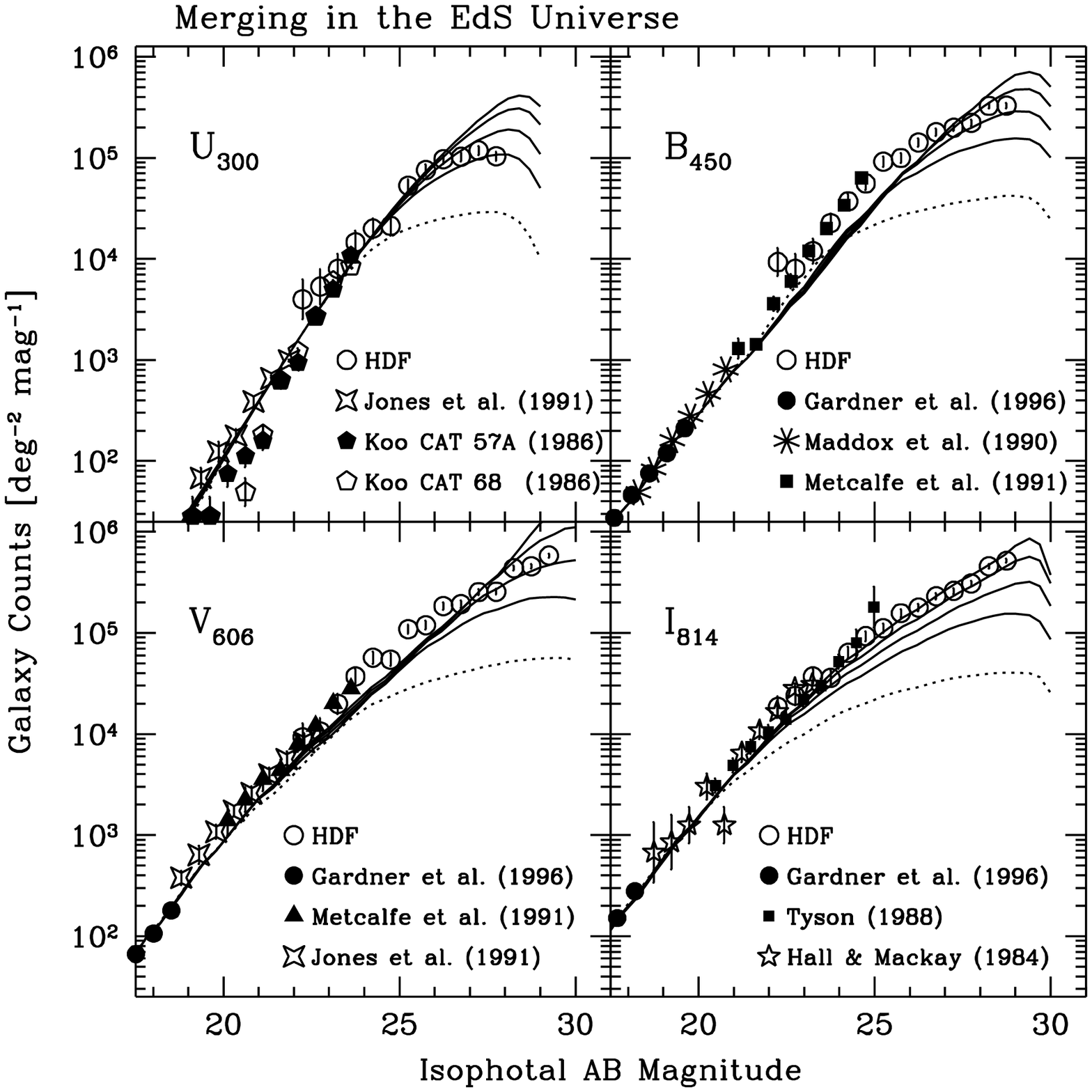,width=13cm}
  \end{center}
\caption{The effect of galaxy number evolution on galaxy counts based 
on the luminosity-density conserving mergers in the Einstein-de Sitter 
(EdS) universe with $(h, \Omega_0, \Omega_\Lambda) = (0.5, 1, 0)$. 
The dotted line is the 
standard PLE model without number evolution in the EdS universe
shown in Fig. \ref{fig:nm-sel-nosel}. 
The four solid lines are the models with the merger parameters of
$\eta =$ 2, 3, 4, and 5 (from the bottom to the 
top), where $\xi$ is fixed to 3.
}
\label{fig:nm-EdS-merge}
\end{figure}

Next we consider the EdS universe where the PLE count prediction is 
by more than one order of magnitudes smaller than those observed
in the HDF (Fig. \ref{fig:nm-EdS-merge}).  The predictions with 
$\eta$ = 2, 3, 4, and 5 are shown by four solid lines in order from
bottom to top, with a fixed value of $\xi = 3$.  The dotted line is 
the PLE prediction without number evolution.  This result indicates 
that a strong number evolution with $\eta \gtilde$ 3--4 is necessary 
to explain the HDF counts in the EdS universe.  We note that, while 
the strong number evolution explains the counts at the faintest 
magnitudes, it fails to explain the overall shape or slope of the 
$N$-$m$ relation.  This failure is clearly seen in the $B_{450}$ 
band, where the strong number evolution makes the count slope less 
steep and deviate from the observations most prominently at 
$B_{450}$ = 22--26.  This argument is quite robust, 
because there should be an upper bound in the steepness of the 
$N$-$m$ slope ($d\log N/dm<$ 0.4), when the total luminosity density of 
all galaxies is conserved during the merger process.  In a more realistic 
case such as gas-rich mergers inducing starbursts, an even flatter 
slope is predicted, because galaxies before mergers are always 
fainter than those in the case of luminosity-density conservation.  
If one tries to explain the observed steep slope by mergers, 
it is necessary to contribe a merger process where galaxies 
before mergers are always brighter, in other words, the luminosity 
density of galaxies increases if the merger process is traced 
backwards---which we consider quite unrealistic.  Therefore, we 
conclude that a strong number evolution in the EdS universe is 
unlikely to explain the observed counts over the whole range of
apparent magnitudes.

\begin{figure}
  \begin{center}
    \leavevmode\psfig{file=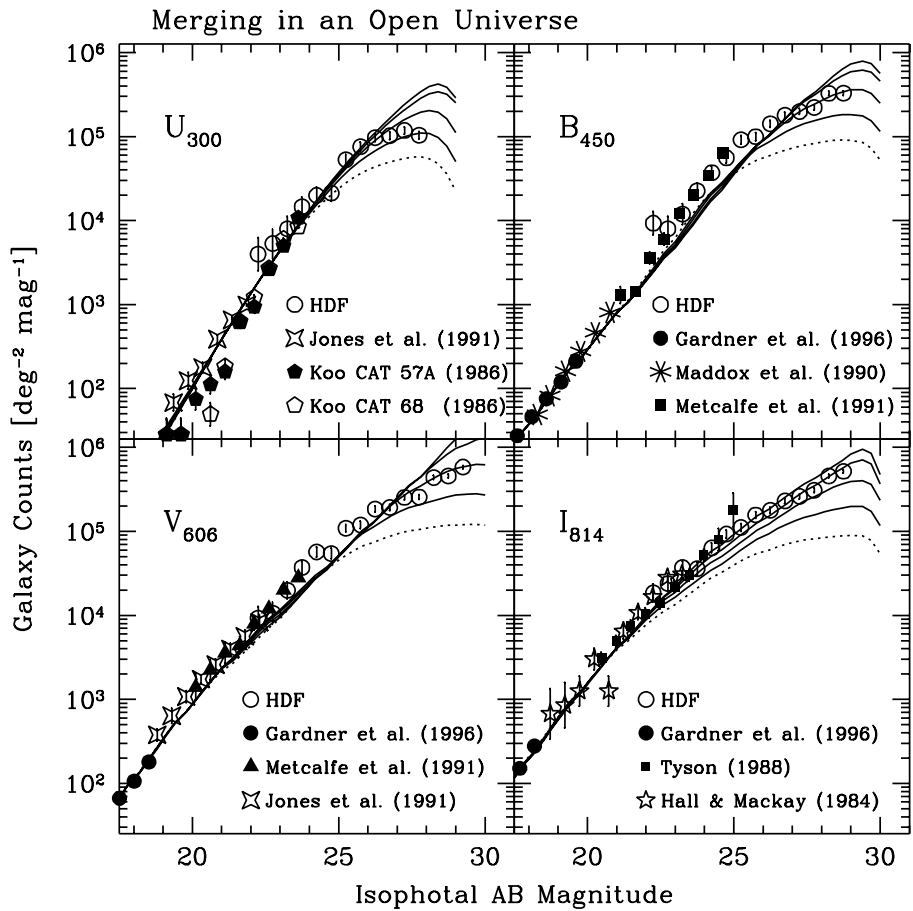,width=13cm}
  \end{center}
\caption{The effect of galaxy number evolution on galaxy counts based 
on luminosity-density conserving mergers in an open
universe with $(h, \Omega_0, \Omega_\Lambda) = (0.6, 0.2, 0)$. 
The dotted line is the 
standard PLE model without number evolution in an open universe
shown in Fig. \ref{fig:nm-sel-nosel}. 
The four solid lines are the models with the merger parameters of
$\eta =$ 1, 2, 3, and 4 (from the bottom to the 
top), where $\xi$ is fixed to 3.
}
\label{fig:nm-open-merge}
\end{figure}

Figure \ref{fig:nm-open-merge} shows the effect of introducing a number 
evolution model in an open universe with $\Omega_0 = 0.2$.  The lines 
in this figure have the same meanings as in Fig. \ref{fig:nm-EdS-merge}, 
but the four solid lines are the predictions with $\eta$ = 1, 2, 3, 
and 4 in order from bottom to top.  In this open universe, a number 
evolution with $\eta \gtilde 2$ is necessary to explain the faintest 
counts, but again it can not explain the steep slope of the $N$-$m$ 
relation in the $B_{450}$ and $V_{606}$ bands.  This indicates that an 
open universe is also difficult to explain the HDF counts if $\Omega_0 > 
0.2$.  However, a lower-density open universe, for example, with $\Omega 
\sim 0.1$ might give a similar result with a $\Lambda$-dominated flat 
universe with $\Omega_0 = 0.2$ (see Fig. \ref{fig:nm-sel-nosel}), and 
such an open universe could also explain the HDF counts if a modest 
number evolution of galaxies is taken into account.

\section{Photometric Redshift Distribution}
\label{section:redshifts}
Although not as reliable as spectroscopic redshifts, photometric 
redshifts of galaxies are useful for statistical studies of 
high-redshift galaxies.  Several groups have published catalogs
of photometric redshifts for the HDF galaxies (e.g., Sawicki, 
Lin, \& Yee 1997; Wang, Bahcall, \& Turner 1998; Fern\'andez-Soto, 
Lanzetta, \& Yahil 1999).  Here we compare our theoretical model with 
the photometric redshift distributions reaching $I_{814} = 28$ derived 
by Fern\'andez-Soto et al. (1999), which utilizes not only the optical 
photometry of the HDF but also the information of the near-infrared
$J$, $H$, and $K$ bands.  We have also compared our model with the
other two catalogs of photometric redshifts by Sawicki et al. (1997) 
and Wang et al. (1998), and confirmed that the following result is 
hardly changed.

\begin{figure}
  \begin{center}
    \leavevmode\psfig{file=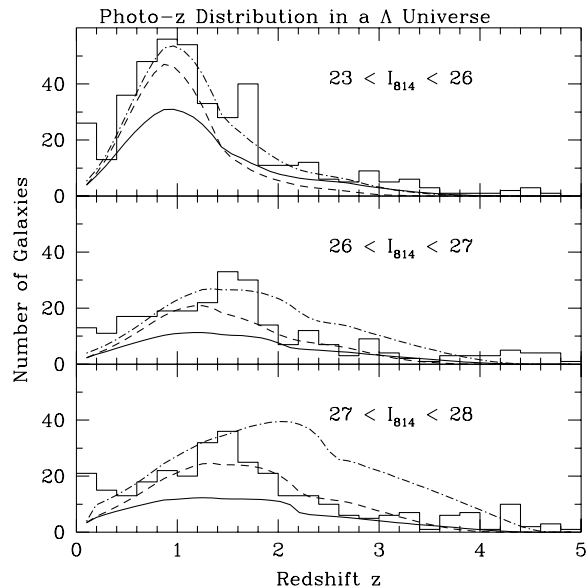,width=8cm}
  \end{center}
\caption{The comparison with the photometric redshift distribution
derived by Fern\'andez-Soto, Lanzetta, \& Yahil (1999) in
a $\Lambda$-dominated flat universe with 
$(h, \Omega_0, \Omega_\Lambda) = (0.7, 0.2, 0.8)$. The solid
line is the standard PLE model without number evolution, with
the same model parameters as in Fig. \ref{fig:nm-type}. The dashed 
line is the merger model with the merger parameters of  $(\eta, \xi)=(1, 3)$, 
which reproduces the observed HDF number counts as well.
The dot-dashed line is the same with the dashed line, but
the selection effects are not taken into account.
}
\label{fig:nz-lam}
\end{figure}

Figure \ref{fig:nz-lam} shows the observed redshift distribution in
three $I_{814}$ magnitude ranges.  The model curves are our predicted 
redshift distributions of galaxies in a $\Lambda$-dominated flat universe 
with $\Omega_\Lambda = 0.8$. The area over which the models are 
calculated is chosen to coincide with the sky area covered by the
observational analysis of Fern\'andez-Soto et al. (1999):
5.31 arcmin$^2$ for $I_{814}<26$ and 3.92 arcmin$^2$ for $I_{814}>26$.  
Therefore,
not only the shape but also the normalization of the predicted redshift 
distributions can be compared directly with the data.
The solid and dashed lines are the models with $(\eta, \xi)=(0,3)$
and (1, 3), respectively, with the selection effects taken into account.  
The dot-dashed line is the same as the dashed line with $(\eta, \xi)=(1,3)$, 
but no selection effects are taken into account.  It is clear that the 
selection effects give a bias against high-redshift galaxies, and this 
selection bias is significant especially at the faintest magnitudes. 
It is inevitable to include these effects when one uses the redshift 
distribution as a probe of number evolution of galaxies.  Comparison 
with the data shows that a modest merger model with $\eta \sim 1$ in 
a $\Lambda$-dominated flat universe gives a reasonable fit to the 
photometric redshift distribution as well as galaxy counts,
provided that the selection effects are properly taken into account. 

\begin{figure}
  \begin{center}
    \leavevmode\psfig{file=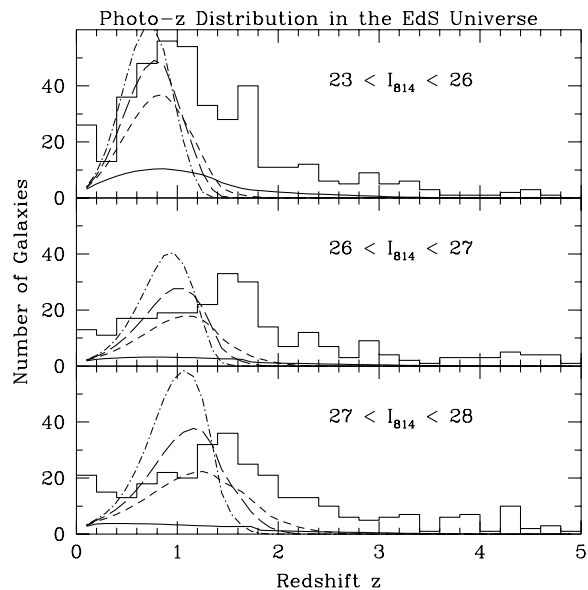,width=8cm}
  \end{center}
\caption{The comparison with the photometric redshift distributions
derived by Fern\'andez-Soto, Lanzetta, \& Yahil (1999) in
the Einstein-de Sitter 
universe with $(h, \Omega_0, \Omega_\Lambda) = (0.5, 1, 0)$. The solid
line is the PLE model without number evolution.
The short-dashed, long-dashed, and dot-dashed lines are the merger
models with $\eta $ = 3, 4, and 5, respectively, where
$\xi$ is fixed to 3. The selection effects are taken into account
in all these models.
}
\label{fig:nz-EdS}
\end{figure}

Figure \ref{fig:nz-EdS} compares the observed redshift distribution 
with the predictions in the EdS universe. The solid line is the 
prediction without number evolution, while other lines are those with 
$\eta$ = 3, 4, and 5 with a fixed value of $\xi$=3.  The selection
effects are taken into account in all curves. A strong number
evolution predicts that most of galaxies have lower redshifts of $z 
\ltilde$ 1, deviating significantly from the observed distribution
for $23 < I_{814} < 26$.  If the assumption of conserved luminosity 
density is relaxed, the predicted distribution becomes peaked
at an even lower redshift, because pre-merger galaxies in more realistic 
merger models are fainter than expected from the conserved luminosity 
density, as discussed in the previous section.  This gives another 
argument that the EdS universe can not explain the observed number of 
HDF galaxies even if a strong merger is invoked, in addition to the 
argument in the previous section
against the EdS universe based on the slope of the observed $N$-$m$ relation.  

\begin{figure}
  \begin{center}
    \leavevmode\psfig{file=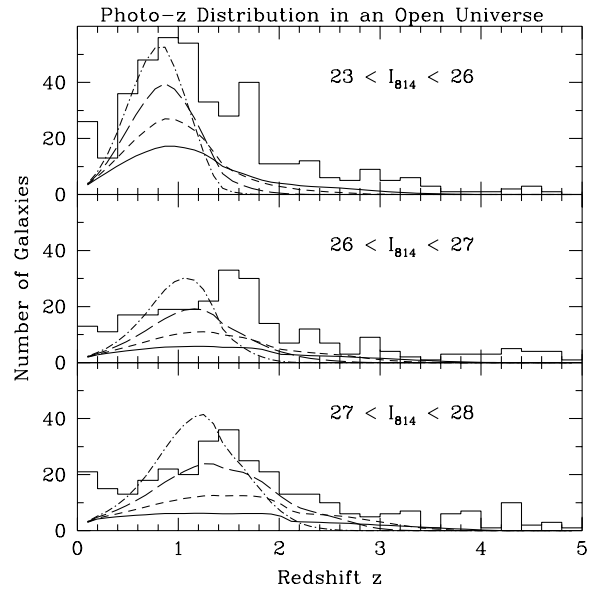,width=8cm}
  \end{center}
\caption{The comparison with the photometric redshift distributions
derived by Fern\'andez-Soto, Lanzetta, \& Yahil (1999). The same as
Fig. \ref{fig:nz-EdS}, but for an open universe with $(h, \Omega_0, 
\Omega_\Lambda) = (0.6, 0.2, 0)$. The solid
line is the PLE model without number evolution.
The short-dashed, long-dashed, and dot-dashed lines are the merger
models with $\eta $ = 1, 2, and 3, respectively, where
$\xi$ is fixed to 3. The selection effects are taken into account
in all these models.
}
\label{fig:nz-open}
\end{figure}

Figure \ref{fig:nz-open} is similar to Fig. \ref{fig:nz-EdS}, but for 
an open universe with $\Omega_0 = 0.2$.  The predicted redshift 
distribution is still peaked at a lower redshift compared with the 
observed distribution.  This discrepancy is significant for
$23 < I_{814} < 26$, but not as serious as in the EdS universe.

\section{Discussion}
\label{section:discussion}

In this paper we have shown that a strong number evolution with 
$\eta \gtilde$ 3--4 is necessary to explain the HDF counts at the 
faintest magnitudes in the EdS universe.  The photometric redshift 
distribution of HDF galaxies suggests that a significant number of 
the faintest galaxies are at $z<1$, and hence there must be a strong 
number evolution already operated at $z<1$ to increase the number of galaxies 
by factor of about 10. Therefore, further argument for or against 
the EdS universe with strong number evolution of galaxies can be made 
from the observational constraints based on spectroscopic redshift 
surveys at $z<1$. 

Totani \& Yoshii (1998) argued that such a strong number evolution at 
$z<1$ is clearly inconsistent with the spectroscopic catalogs of 
galaxies reaching $z \sim 1$, at least for giant galaxies with $L 
\sim L^*$.  This is based on a $V/V_{\max}$ test for the galaxies 
in the Canada-France Redshift Survey (CFRS, Lilly et al. 1995), 
and an important finding is that the PLE model is not inconsistent with the 
spectroscopic data, giving a constraint on number evolution as 
$\eta = 1.8 \pm 0.7$, $1.1 \pm 0.7$, and $0.5 \pm 0.7$ for the EdS 
universe, an open $\Omega_0=0.2$ universe, and 
a flat $\Omega_\Lambda=0.8$ universe,
respectively.  Although the $V/V_{\max}$ test favors a larger $\eta$ 
in the EdS universe, this value seems smaller than that required to 
explain the HDF counts at the faintest magnitudes.  These results 
have been confirmed for elliptical galaxies by Shade et al. (1999), 
in which they found that the population of massive early-type galaxies 
was largely in place by $z \sim 1$.  

An independent constraint on galaxy mergers at $z < 1$ comes from 
statistical studies of merging galaxies inferred from high-resolution 
images.  Recently, Le F\'evre et al. (2000) derived the evolution of 
merger rate from the HST images of the CFRS and LDSS galaxies.  Their 
result suggests that $L^*$ galaxies on the average have undergone 
about one merger event from $z = 1$ to 0, which corresponds to $\eta 
\sim 1$ and hence this is consistent with the number evolution in a 
$\Lambda$-dominated flat universe suggested by this paper.

Pozzetti et al. (1998) claimed that the PLE model can not explain all 
the observed data, although it well explains the HDF galaxy counts.
The major discrepancies between the PLE model and the observed data
were found in the evolution of the luminosity density in the universe.
That is, the observed luminosity density increases more steeply to 
$z\sim 1$ than the PLE model prediction, 
and on the other hand, the PLE model predicts too high
UV luminosity density at $3.5 < z < 4.5$ compared
with the observation, because of intense starbursts in elliptical galaxies.
However, Totani, Yoshii, \& Sato (1997) had 
already pointed out that the observed steep evolution to $z \sim 1$ is 
explained in a $\Lambda$-dominated flat universe.
It has already been 
argued that the PLE model is consistent with the observed luminosity 
density evolution if initial starbursts in high-redshift elliptical 
galaxies are obscured or not existent.  In fact, our calculation of 
the redshift distribution in this paper shows that initial starbursts
are not detected at $z \gtilde 3$, because of the dust obscuration and 
the selection effects.  Therefore, the major problems of the PLE model 
claimed by Pozzetti et al. (1998) are resolved, and the PLE model in a 
$\Lambda$-dominated flat universe gives a reasonable fit to the observed 
HDF data, allowing only for a modest number evolution of galaxies with 
$\eta \ltilde 1$.

\section{Summary and Conclusions}
\label{section:conclusions}
We have modeled the number count and redshift distribution of faint 
HDF galaxies, with the observational selection effects properly taken 
into account.  As a consequence of the selection effects in the 
theoretical modeling, predicted counts from the PLE model are
smaller than previously considered, and they are more than 10 
times smaller than the observed HDF counts at the faintest magnitudes 
in the EdS universe.  A strong number evolution with $\eta \gtilde$ 
3--4 under the assumption of conserved luminosity density is
required to explain the faintest counts in this EdS universe, 
when the number evolution is
parametrized as $\phi^* \propto (1+z)^\eta$ and 
$L^* \propto (1+z)^{-\eta}$.  However, such a strong number evolution 
is not consistent with the overall $N$-$m$ slope or the photometric 
redshift distribution.  These discrepancies become even worse when one 
considers a more realistic merger process, i.e., enhanced star formation 
following by gas-rich mergers.  In addition, such a strong evolution is 
rejected at least for average $L^*$ galaxies at $z<1$ from the data of 
spectroscopic redshift surveys (Totani \& Yoshii 1998;
Shade et al. 1999; Le F\'evre et al. 2000).  Therefore, we 
conclude that it is almost impossible to explain the HDF galaxies in 
the EdS universe, unless we invoke ultra-exotic galaxy populations such 
as galaxies forming only massive stars at high redshifts to escape from 
local galaxy surveys due to the complete lack of long-lived stars.  

The present work revitalizes the practice of using faint number counts
as an important cosmological test,
which gives one of the arguments against the EdS universe 
by its outstanding statistics compared with other cosmological tests. 
An open universe with $\Omega_0 > 0.2$ does not fit to the HDF data either,
for the similar reasons for rejecting the EdS universe.  An open 
universe with $\Omega_0 \sim 0.1$ might be consistent with the HDF 
data, but such a low value of $\Omega_0 \sim 0.1$ would not be reconciled 
with other constraints on $\Omega_0$, such as the baryon-gas to 
dark-matter mass ratio in poor clusters of galaxies combined with the 
standard big-bang nucleosynthesis prediction of baryon mass density in 
the universe (e.g., Pedersen, Yoshii, \& Sommer-Larsen 1997). 

We have extensively 
checked systematic uncertainties in our theoretical modeling
of galaxy formation and evolution, and found that they 
are unlikely to resolve the above discrepancies emerged in the EdS 
universe and also in an open universe.  On the other hand, such 
discrepancies are naturally resolved if we invoke a $\Lambda$-dominated 
flat universe.  This suggests that the existence of the cosmological 
constant or an exotic form of the vacuum energy density of the universe 
which is now accelerating the expansion of the universe. 

The PLE model in a $\Lambda$-dominated flat universe with $\Omega_0 
\sim 0.2$ gives a reasonable fit to the HDF data, and a modest number 
evolution with $\eta \ltilde 1$ is also suggested by the HDF counts at 
the faintest magnitudes.  It should be noted that this number evolution 
does not necessarily mean mergers of galaxies, but may suggest strongly 
clumpy star-forming regions within an individual galaxy system becoming 
visible at high redshifts (Colley et al. 1996; Bunker, Spinrad, \& 
Thompson 1999).

On the other hand, it is interesting to note that this indication of
mild number evolution is consistent with the merger rate evolution of 
$L^*$ galaxies at $z<1$ recently inferred from a high-resolution 
image study for galaxies in the CFRS survey (Le F\'evre et al. 2000). 
This result is consistent with some models of galaxy formation 
based on the hierarchical structure formation in the CDM universe 
(Le F\'evre et al. 2000), although there are considerable uncertainties 
in the theoretical calculations for the merging history of baryonic 
component.  A stronger number evolution with $\eta \gtilde 1$ is, 
however, strongly disfavored by the observed HDF galaxy counts.  This will 
give an important constraint when galaxy formation is modeled in the 
framework of the structure formation in a cold dark matter universe.

Inclusion of the selection effects in this paper leads to a considerably 
different result from previous studies on galaxy number count and redshift 
distribution.  This means that any cosmological interpretations will be 
seriously misled if the selection effects are ignored.  All future studies 
related to the detection and statistics of high-redshift galaxies should 
take into account these effects.  The selection effects give a bias against 
high-redshift galaxies, reducing a problem of overprediction of such galaxies 
by the PLE model, which has been claimed by several studies ignoring the 
selection effects (e.g., Ellis 1997).  
In fact, we have shown that the PLE model is in overall
agreement with the HDF galaxies, even if a modest number evolution of galaxies 
($\eta \ltilde 1$) may be required.  A strong number evolution, however, 
predicts too small a number of high-redshift galaxies to be consistent with 
the photometric redshift distribution of the HDF galaxies.

The authors would like to thank K. Shimasaku for providing numerical 
data for the filter functions of HST photometry bands, 
and T. Tsujimoto and C. Kobayashi for providing their models of
galaxy evolution in a tabular form.  We also thank an anonymous referee
for many useful comments which have considerably improved this manuscript.
This work has been supported in part by 
a Grand-in-Aid for Conter-of-Excellence Research (07CE2002) of the Ministry 
of Education, Science, and Culture in Japan.


\newpage


\begin{table}
\caption{Schechter Parameters Adopted for Local Luminosity Functions}
\begin{tabular}{lcccccc}
\hline
\hline
Survey & Type & $\phi^*$ [Mpc$^{-3}$] & $\alpha$ & $M_B^*$ & References \\
\hline
SSRS2 & E/S0 & $4.4\times 10^{-3}$ & $-1.00$ & $-19.37$ & 1 \\
      & Sab--Scd & $8.0\times 10^{-3}$ & $-1.11$ & $-19.43$ \\
      &  Sdm     & $0.2\times 10^{-3}$ & $-1.81$ & $-19.78$ \\
\hline
Stromlo-APM & All & $1.4 \times 10^{-2}$ & $-0.97$ & $-19.50$ & 2\\
\hline
CfA   & E/S0 & $4.6 \times 10^{-3}$ & $-0.88$ & $-19.23$ & 3\\
      & Sab  & $3.8 \times 10^{-3}$ & $-1.00$ & $-18.99$ \\
      & Sbc  & $6.6 \times 10^{-3}$ & $-0.44$ & $-19.06$ \\
      & Scd  & $2.0 \times 10^{-3}$ & $-1.00$ & $-19.34$ \\
      & Sdm  & $0.8 \times 10^{-3}$ & $-1.20$ & $-19.29$ \\
\hline
\hline
\end{tabular}

NOTE.---$H_0 = 100$ km/s/Mpc. SSRS2 is adopted as a
standard in our analysis. The value of $M_B^*$ is given for the $B$
band, which is related to $B_{450}$ as $B_{450} = B - 0.153$.\\
REFERENCES.---(1) Marzke et al. (1998), (2) Loveday et
al. (1992), (3) Huchra et al. (1983). See also Efstathiou, Ellis,
\& Peterson (1988) and Yoshii \& Takahara (1989).
\label{table:lf}
\end{table}

\begin{table}
\caption{Figure Identification for  Galaxy Count Calculations}
\begin{tabular}{lc}
\hline \hline
Examined Effects & Figure No. \\
\hline
Galaxy morphological types   & \ref{fig:nm-type} \\
Selection effects  & \ref{fig:nm-sel-nosel} \\
Cosmological parameters & \ref{fig:nm-sel-nosel} \\
Galaxy luminosity evolution & \ref{fig:nm-model} \\
Absorption (dust and intergalactic HI clouds) & \ref{fig:nm-model} \\
Local luminosity function & \ref{fig:nm-zFlf} \\
Formation redshift $z_F$ & \ref{fig:nm-zFlf} \\
Dispersion in the $L_B$-$r_e$ relation & \ref{fig:nm-r-e} \\
Merging of galaxies & \ref{fig:nm-lam-merge}--\ref{fig:nm-open-merge} \\
\hline \hline
\end{tabular}
\label{table:nm-summary}
\end{table}

\begin{table}
\caption{Sensitivity of the Predicted $N$-$m$ Relation to 
the Change of Input Parameters}
\scriptsize
\begin{tabular}{clccccccccccc}
\hline \hline 
&& \multicolumn{2}{c}{$U_{300}$}  &
& \multicolumn{2}{c}{$B_{450}$}  &
& \multicolumn{2}{c}{$V_{606}$}  &
& \multicolumn{2}{c}{$I_{814}$}  \\
\cline{3-4} \cline{6-7} \cline{9-10} \cline{12-13}
\# & Change 
& $\Delta \log N_{\rm 25 mag}$ 
& $\Delta \log N_{\rm 28 mag}$ &
& $\Delta \log N_{\rm 25 mag}$ 
& $\Delta \log N_{\rm 28 mag}$ &
& $\Delta \log N_{\rm 25 mag}$ 
& $\Delta \log N_{\rm 28 mag}$ &
& $\Delta \log N_{\rm 25 mag}$ 
& $\Delta \log N_{\rm 28 mag}$ \\
\hline
1& Cosmology: $\Lambda \rightarrow$ EdS/open$^a$ & -0.227/-0.126 
& -0.513/-0.232 && -0.446/-0.235 & -0.610/-0.278
&& -0.493/-0.254 & -0.601/-0.258 && -0.522/-0.268 & -0.614/-0.284 \\
2& Evolution: AY, AYT$\rightarrow$KTN$^b$ & +0.048 & -0.026 && -0.027 & -0.086 
&& -0.056 & -0.017 && -0.044 & +0.006 \\
3& Dust absorption: screen$\rightarrow$slab 
& -0.085 & -0.166 && +0.062 & +0.101 && +0.137 & +0.136 && +0.180 & +0.153 \\
4& HI absorption: on$\rightarrow$off & +0.094 & +0.138 && +0.020 & +0.009 
&& +0.005 & +0.008 && 0.000 & 0.000 \\
5& Local LF: SSRS2$\rightarrow$Stromlo-APM/CfA$^c$ 
& -0.119/+0.066 & -0.283/-0.225 && -0.071/+0.041 & -0.218/-0.128
&& -0.042/+0.022 & -0.187/-0.168 && -0.055/-0.007 & -0.183/-0.170 \\
6& Formation epoch $z_F$: $5\rightarrow$ 3/10 & +0.054/-0.040 & -0.101/+0.036 
&& -0.070/-0.062 & -0.151/+0.075 && -0.119/-0.025 & -0.151/+0.079 
&& -0.110/+0.017 & -0.154/+0.091 \\
7& $r_e$-$L_B$ relation: $+1\sigma$/$-1\sigma$ in $\Delta(\log r_e)^d$ 
& -0.088/+0.045 & -0.159/+0.103
&& -0.150/+0.069 & -0.160/+0.098 && -0.104/+0.050 & -0.147/+0.094
&& -0.113/+0.061 & -0.167/+0.106 \\
\hline
 & Total systematic uncertainty$^e$ & $\pm$0.157 & $\pm$0.339 
&& $\pm$0.155 & $\pm$0.277 && $\pm$0.185 & $\pm$0.279 &&
$\pm$0.217 & $\pm$0.298  \\
\hline \hline
\end{tabular}
$^a\Lambda$: $(\Omega_0, \Omega_\Lambda)$ = (0.2, 0.8), EdS: (1, 0),
open: (0.2, 0) \\
$^b$AY: Arimoto \& Yoshii (1987), AYT: Arimoto, Yoshii, \& Takahara
(1992), KTN: Kobayashi, Tsujimoto, \& Nomoto (1999).\\
$^c$See Table \ref{table:lf} for detail. \\
$^d$See \S \ref{section:size-luminosity} for detail. \\
$^e$Quadratic sum of the rows 2, 3, 5--7. A mean value of the two numbers
in the rows 5--7 is used in the sum. \\
NOTE.---The prescriptions of a standard model in our analysis include
a $\Lambda$ cosmology, AY-AYT evolution model, the local LF of SSRS2,
$z_F=5$, the screen model of dust absorption, and the intergalactic
HI absorption. \\
\vspace*{0.3cm}
[Complete version of this table is available at 
http://th.nao.ac.jp/~totani/images/paper/ty2000-table3.ps.]

\label{table:nm-sensitivity}
\end{table}


\end{document}